\documentclass[%
 reprint,
 aps,
]{revtex4-2}
\usepackage{tikz}
\usetikzlibrary{positioning}
\usepackage{amsmath}
\usepackage{amsthm}
\usepackage{float}
\usepackage{braket}
\usepackage{graphicx}
\usepackage{braket}
\usepackage{cancel}
\usepackage{bm}
\usepackage{qcircuit}
\usepackage{url}
\usepackage{hyperref}
\usepackage{simplewick}

\newtheorem{thm}{Theorem}[section]
\newtheorem*{thm*}{Theorem}
\newtheorem{defn}[thm]{Definition}
\newtheorem*{defn*}{Definition}

\newtheorem*{lem*}{Lemma}

\newtheorem*{rem*}{Remark}

\newtheorem*{con*}{Conjecture}

\newtheorem*{cor*}{Corollary}
\newtheorem{prop}[thm]{Proposition}
\newtheorem*{prop*}{Proposition}

\newtheorem*{hypoth*}{Hypothesis}

\newtheorem*{claim*}{Claim}
\newtheorem*{prf}{Proof}

\begin{document}
\preprint{APS/123-QED}

\title{Theory of Quantum Games and Quantum Economic Behavior}

\author{Kazuki Ikeda}\email{kazuki7131@gmail.com}
\affiliation{Department of Mathematics and Statistics
$\&$ Centre for Quantum Topology and Its Applications (quanTA), University of Saskatchewan, Saskatoon, Saskatchewan S7N 5E6, Canada}

\author{Shoto Aoki}
\affiliation{Department of Physics, Osaka University, Toyonaka, Osaka 5600043, Japan }

\date{November 11, 2021}

\begin{abstract}
The quest of this work is to present discussions of some fundamental questions of economics in the era of quantum technology, which require a treatment different from economics studied thus far in the literature. A study of quantum economic behavior will become the center of attention of economists in the coming decades. We analyze a quantum economy in which players produce and consume quantum goods. They meet randomly and barter with neighbors bilaterally for quantum goods they produced. We clarify the conditions where certain quantum goods emerge endogenously as media of exchange, called quantum commodity money. As quantum strategies are entangled, we find distinctive aspects of quantum games that cannot be explained by conventional classical games. In some situations a quantum player can acquire a quantum good from people regardless of their strategies, while on the other hand people can find quantum strategies that improve their welfare based on an agreement. Those novel properties imply that quantum games also shed new light on theories of mechanism design, auction and contract in the quantum era.  
\end{abstract}

\maketitle

\onecolumngrid
\section{Introduction}
\subsection{General Remark}
Recent progress in quantum technologies has been stood us in good stead to acquire abilities to control quantum systems for various usages. It runs the gamut from quantum computing \cite{1982IJTP...21..467F,doi:10.1098/rspa.1985.0070} to communication with quantum network \cite{Elliott_2002,10.1117/12.606489,Peev_2009,Sasaki:11,dynes2019cambridge}. Those quantum technologies are bound to have a significant impact on economy in the literature. In particular the appearance of programmable quantum computers \cite{arute2019quantum,debnath2016demonstration} heralds potential growth in demand for goods or services available via the quantum internet \cite{kimble2008quantum}, by interconnecting quantum computers with the quantum teleportation between nodes \cite{PhysRevLett.70.1895,1997Natur.390..575B,Furusawa706,2013Natur.500..315T}. {In addition, various quantum communication channels for realizing genuine quantum-mechanical communication are being explored and implemented~\cite{PhysRevA.92.012317, 2010, Adnane:19, PhysRevA.93.062315}. } With quantum communication devices and networks, economic activities will be buoyed in conformity with the laws of quantum physics. Consequently they promise the rise of quantum economics, which is a branch of economics in the era of noisy intermediatescale quantum (NISQ) technologies \cite{preskill2018quantum} and full-fledged quantum technologies in future \cite{aaronson2011computational,preskill2012quantum}. 

Quantum game theory is a theoretical study on efficient and rational ways of programming quantum devices. It will accrue to multifarious activities of humanbeings. In this article we aim at providing a theoretical study on the quantum economical behavior in a newly emerging quantum information and communication technology (QICT) market. To our best knowledge this is the first paper that expounds on economic activities from a viewpoint of quantum games. Indeed, mathematical or physical aspects of quantum games, such as the role of entanglement or quantum strategies for mathematical toy models, have been investigated in many works \cite{PhysRevLett.83.3077,PhysRevLett.82.1052,2001PhRvA..64c0301B}. However, less is known from an economic perspective that leads to a meaningful analysis of a quantum market doctrine and of economical behavior in the market. Therefore a critical milestone in quantum games that shows a significant difference from the traditional games has not been reached. This unfortunate lack of impact on economics casts doubt on the power of quantum computational protocols for games. The major criticisms toward the present situation of quantum games are summarized in \cite{2002PhRvA..66b4306V,2018arXiv180307919S}. In this article we tackle this challenge and succeed in showing novel key differentiating features of a quantum game model. The classical game theory today \cite{morgenstern1953theory} plays a crucial role in modern economics \cite{aumann1992handbook} and based on its history we can foresee prosperous days of quantum games as the foundation of quantum economics.  

\subsection{Quantum Information as Economic Commodity}
We study economies where quantum commodities that serve as media of exchange are determined endogenously. In the modern economy, almost all products are exchanged for money, which is defined as an object with functions of a medium of exchange, a unit of account and a store of value. In the long period of history of money, it was only as late as the second half of the 20 century that the modern fiat money system set by command of the central bank was established. Precious metals had been widely used as a commodity money till the U.S. government abandoned the convertibility of the dollar into gold. A physical form of money also keeps changing. The recent advances in technologies popularize the use of digital money as an official alternative of legal tender. More recently cryptoassets and cryptocurrencies \cite{nakamoto2019bitcoin} exude presence in various and sundry fields \cite{ZHANG20181,BANERJEE201869,IKEDA201899}. Therefore it is meaningful to investigate the possibility of a system and a form of money in the era of quantum technologies. The concepts of quantum money are proposed multifariously \cite{Wiesner:1983:CC:1008908.1008920,2010arXiv1004.5127F,ikeda2018security,ikeda2018qbitcoin}, however, whether quantum objects can be media of exchange is yet another question. Quantum money is a naive quantum extension of classical money, but there are no formal modelings that show the medium of exchange function as an equilibrium of a quantum game. To proceed our discussion, it is important to recall that people in the modern society indeed exchange, create and consume classical information in many situations of online activities in which they do not directly process macroscopic goods and all of those transactions are based on some bit strings of classical information. Therefore classical information is an economical commodity. In addition, people do not directly consume an electron when they enjoy online shopping, although their information should come from some electrical devices. The same things will be realized by means of quantum devices. 

In this work we show how quantum information can become an economical commodity. {In other words, if there is a certain number of people who want to exchange quantum information through a quantum network, and if there is a certain amount of supply and demand, quantum information itself will have value as an economic commodity. However, some people may feel uncomfortable with the idea of value being created from invisible quantum states. Our answer to this question is clear. First, even classical information is now stored and exchanged electronically and invisibly. The digitized information has the same amount of information as the original, and is actually used in a variety of situations. In terms of cost, it is more convenient to hold information in electronic form. And for those who are accustomed to holding information as digital information, there should not be much resistance to holding information in quantized form. This is because information composed of bits can be embedded directly into qubits. As long as quantum information can be stored in quantum memory and quantum devices can be networked together, we can communicate in a quantum society in the same way that people have done in digital societies. In that case, what is the advantage of holding the system in a quantum state is the most important question. It is known that the use of quantum information increases the confidentiality of the information holder, and that the use of quantum networks significantly increases security. In other words, by quantizing information, it is possible to send and receive extremely important information safely and securely. The important point here is that such important information generally has a high monetary value. For example, a list of buyers of a certain product or personal information is generally bought and sold in a market, even if the possession of the information itself does not immediately lead to money. Therefore, even if the information held as quantum information does not immediately lead to money, it can have sufficient monetary value if it is in demand among certain people and can be traded. This is what is meant by quantum information being an economic commodity. Therefore, even if the information held as quantum information does not immediately lead to money, it can have sufficient monetary value if it is in demand among certain people and can be traded. This is what is meant by quantum information being an economic commodity. And as some of these economic commodities are distributed in the market, some of them will function as commodity money. The main interest of this paper is to analyze clearly and rigorously from a game theoretic point of view what kind of quantum states can become commodity quantum money.
}

\subsection{Summary of Our Contribution}
The main purpose of this article is to analyze a simple quantum model, where quantum objects which become media of exchange are pursed as a theory of non-cooperative repeated quantum game \cite{ikeda_foundation_2019,2020arXiv200505588A}. The medium of exchange function is a crucial property of money. The historical evolution of money has long been studied by many authors. Karl Marx \cite{marx1911contribution} listed the functions of money for hoarding and Adam Smith \cite{smith1950inquiry} described that the idea to overcome the double coincidence of wants \cite{jevons1876money} leads to the use of money as a medium of exchange. The formal modeling of equilibria with classical commodity money and classical fiat money is known as the Kiyotaki-Wright model \cite{10.2307/1832197}, which explores the economic foundation of the emergence of money as a perishable medium of exchange. The model is tested numerically \cite{Duffy99,Duffy98fiatmoney,BROWN1996583,MARIMON1990329} and a mixed strategy case is also proposed \cite{Kehoe93}. Kiyotaki and Wright initially addressed infinitely durable goods and later their model was extended to goods with a different durability \cite{Cuadras-Morato1997, kawagoe}. In this article we purse the conditions and welfare properties of quantum objects in order to clarify how trade using quantum media of exchange can emerge in equilibrium. The main achievement in Sec.\ref{sec:theorem} is Theorem \ref{thm:1}, which clarifies a mechanism of the emergence of money as a medium of exchange. We instantiate some strategies and equilibria in Sec.\ref{sec:entangled_goods}. In Sec.\ref{sec:entangled_strategy} we investigate more details of entangled quantum strategies. A type of quantum goods which can be a medium depends on strategies. Some concrete examples are summarized in Table \ref{tab:strategies}. It is notable that quantum entanglement among strategies allow Bob to obtain his consuming good from Alice regardless of her inclination. This is a distinctive aspect of our quantum economy and not found in classical economy. In other words, those characteristic properties of quantum games incentivize us to investigate theories of mechanism design, auction and contract from a perspective of quantum information theory. We assume all quantum goods are infinitely durable by means of various quantum information processing technologies, such as quantum error correction \cite{PhysRevLett.77.793,1997PhDT.......232G,1996PhRvL..77.3260D,PhysRevA.52.R2493}, fault-tolerant quantum computation \cite{RAUSSENDORF20062242,2007PhRvL..98s0504R,2007NJPh....9..199R} and robust quantum memories \cite{2002JMP....43.4452D,2008PhRvL.100p0501G}. As a generic quantum state has finite lifetime, it will be also interesting to consider quantum goods with a different durability. 

\section{Quantum Money as a Medium of Quantum Exchange}
\subsection{\label{sec:money}Quantum Commodity Money}
 A quantum object is called a quantum commodity money if it is accepted in trade not to be consumed or used in production, but to be used to facilitate further trade. If a quantum object with no intrinsic value becomes a medium of exchange, it is called quantum fiat money. In this article we discuss how a quantum commodity can be a medium of exchange. 

As a simple model, we consider a quantum information system in which there are three commodities called goods 1,2 and 3. Quantum agents in the community are classified into one of three types: type $i$ quantum agent can produce a good $j(\neq i)$ and receive positive utility only from consumption of a good $i$. In model A, type $i = 1, 2, 3$ agents produce goods $j = 2, 3, 1$, respectively, while in model B type $i = 1, 2, 3$ produce a good $j =3, 1, 2$, respectively. At every round, agents who are allowed to
store one unit of any good need to decide whether they carry over their goods to the next round or exchange. In each period, agents are
matched randomly (with the probability 1/3) in pairs and must decide whether or not to trade bilaterally. When type $i$ gets a good $i$, he or she immediately consumes it, produces a new unit of a good, and stores it until the next date. Storing a good is costly, and an agent incurs storage costs $c_{j}$ for holding a good $j$ at the end of every round. When an agent of type $i$ successfully trades for a good $i$, then he immediately consumes that good and produces a new unit of his production good $j$ without cost. All agents receive the same profit $u(>c_k~\forall k)$ from consuming goods. Therefore the total classical profit $\pi_i$ of an agent of type $i$ is given by 
\begin{equation}
\label{eq:profit}
    \pi_i(t)=\sum_{\tau=0}^t (f_i(\tau)u-p_{ij}(\tau)c_{ij}) \delta^{\tau}, 
\end{equation}
where $\delta\in(0,1)$ is a discount factor, $p_{ij}(t)$ is $i$'s probability to have a good $\ket{j}$, $f_i(t)$ is $i$'s probability to get and consume $i$ at date $t$, and $c_{ij}$ is the cost to type $i$ of storing a good $j$. Note that $f_{i}(t)$ depends on strategies and $p_{ij}(t-1)$. 

Agents try to maximize their total profit. \if{The parameters of this study is shown in Table \ref{tab:my_label}.
\begin{table}[h]
    \centering
    \begin{tabular}{c|c|c}
    Parameters &Model A& Model B\\\hline
       $u$  & 20 or 100 & 20 or 500 \\
        $\beta$ & 0.9 &0.9\\
        $c_1$ & 1&1\\
        $c_2$ & 4 & 4\\
        $c_3$ & 9 &9\\ \hline
    \end{tabular}
    \caption{Parameters for this study. The same values given in \cite{10.2307/117162,Duffy:2001} are used for the purpose of comparison. As a repeated quantum game, $\beta\in(0,1)$ is the discount factor common across type.}
    \label{tab:my_label}
\end{table}
}\fi
In both models, there are two equilibria: fundamental equilibrium and speculative equilibrium. One equilibrium is referred to as a fundamental equilibrium, which is dominant in model A, since agents always prefer a lower-storage-cost commodity to a higher-storage-cost commodity unless the latter is their own consumption good. On the other hand, in speculative equilibrium, which is dominant in model B, agents trade a lower for a higher-storage-cost commodity not because they wish to consume it, but because they rationally expect that this is the best way to ultimately trade for another good that they do want to consume, that is, because it is more marketable. 

In this article, we address the type A model and assume there are only three persons Alice, Bob and Charlie who represent type 1,2 and 3 agents respectively. So, Alice creates a type 2 quantum good $\ket{2}$ and consumes a type 1 quantum good $\ket{1}$. Bob creates a type 3 quantum good $\ket{3}$ and consumes a type 2 quantum good $\ket{2}$. Charlie creates a type 1 quantum good $\Ket{1} $and consumes a type 3 quantum good $\ket{3}$. For example, Alice gives Bob a type $2$ or $3$ quantum good to get a type $1$ quantum good (Fig.\ref{fig:exchange}). 
\if{
\begin{figure}[H]
    \centering
\resizebox{0.5\columnwidth}{!}{%
\begin{tikzpicture}[auto,->]
\node (A) {Alice};
    \node[below left=of A] (B) {Bob};
    \node[below right=of A] (C) {Charlie};
\draw[transform canvas={yshift=2pt}]
(A) -- node[swap] {$\scriptstyle 2,3$} (B);
\draw[transform canvas={yshift=-2pt}]
(B) -- node[swap] {$\scriptstyle 1,3$} (A);
\draw[transform canvas={yshift=2pt}]
(A) -- node[swap] {$\scriptstyle 1,2$} (C);
\draw[transform canvas={yshift=-2pt}]
(C) -- node[swap] {$\scriptstyle 2,3$} (A);
\draw[transform canvas={yshift=2pt}]
(B) -- node {$\scriptstyle 1,3$} (C);
\draw[transform canvas={yshift=-2pt}]
(C) -- node[below] {$\scriptstyle 2,3$} (B);
\end{tikzpicture}  
}
    \caption{The diagram of trades.}
    \label{fig:exchange}
\end{figure}
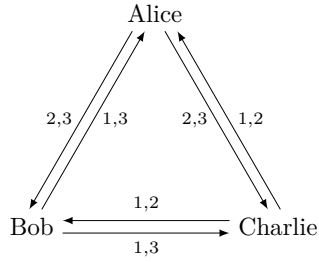
}\fi

\begin{figure}[H]
    \centering
    \includegraphics[width=5cm, bb=0 0 150 100]{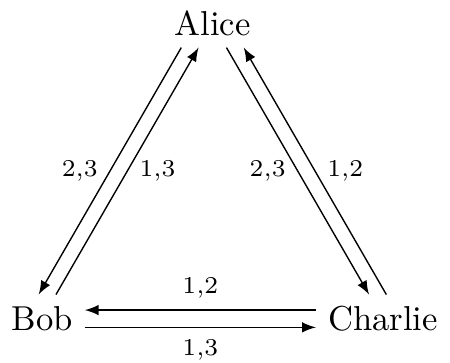}
\caption{The diagram of trades.}
    \label{fig:exchange}
\end{figure}

\subsection{\label{sec:circuit}Quantum Circuit for Consensus Building}
\begin{figure}[H]
\centering
\if{
\begin{quantikz}
\lstick{$\ket{\psi_A}$} & \gate[2]{\mathcal{K}} & \phase{0}\vqw{2} &\qw&\qw&  \swap{2}  &\qw  \\
\lstick{$\ket{\psi_B}$} &&\qw & \phase{1}\vqw{2} &\qw&\targX{}  &\qw \\
\lstick{$\ket{0}_A$} & \gate[2]{\mathcal{J}}  & \gate{U_A}&  \qw & \gate[2]{\mathcal{J}^\dagger} &\phase{1}\vqw{-1}& \meter{}&\\
\lstick{$\ket{0}_B$} & &\qw& \gate{U_B}  && \phase{1}\vqw{-1}& \meter{} &   
\end{quantikz}}\fi
    \includegraphics[width=6cm, bb=0 0 200 100]{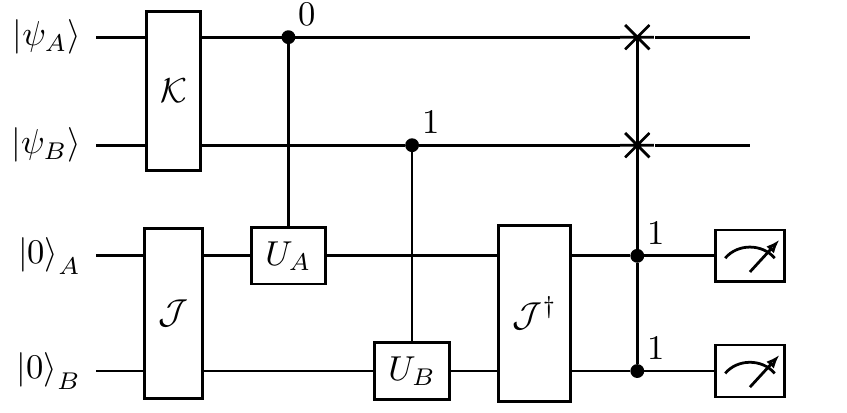}
    \caption{Quantum circuit in which Alice and Bob exchange her $\ket{0}$ and his $\ket{1}$, when they have. $\mathcal{K}$ creates entanglement between states $\ket{\psi_A}$ and $\ket{\psi_B}$, and $\mathcal{J}$ yields entanglement between quantum strategies $U_A$ and $U_B$. $\mathcal{J}^\dagger$ is operated so the game recovers classical results when $U_A$ and $U_B$ are classical strategies. The swap operator acts on the first two qubits if two agents agree to trade.}
    \label{fig:two_persons}
\end{figure}

As the most fundamental case of consensus building, we first describe a trade between two persons. To clarify our idea, we exhibit a quantum circuit (Fig.\ref{fig:two_persons}) in which Alice and Bob exchange her $\ket{0}$ and his $\ket{1}$. They operate their unitary matrices to their ancilla $\ket{0}_A$ and $\ket{0}_B$ in such a way that 
\begin{align}
\begin{aligned}
    U_A\ket{0}_A&=a_0\ket{0}_A+a_1\ket{1}_A=:\Ket{\tau_A (0, 1)} \\
    U_B\ket{0}_B&=b_0\ket{0}_B+b_1\ket{1}_B=:\Ket{\tau_B (1, 0)},  
\end{aligned}
\end{align}
where we interpret $|a_1|^2~(|b_1|^2)$ as Alice's (Bob's) probability of willing to trade. Their deal can be made if and only if $\ket{1}_A\ket{1}_B$ is observed in the ancilla state. 

\begin{figure}[H]
\centering
\if{
\resizebox{0.2\columnwidth}{!}{%
\begin{quantikz}
  & \phase{i}\vqw{1} &\qw& \\
 & \gate{U}&\qw& 
\end{quantikz}
}}\fi
\includegraphics[width=3cm, bb=0 0 50 50]{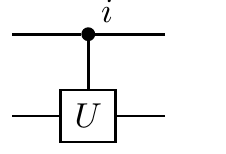}
    \caption{Quantum gate for an exchange operation.}
    \label{fig:control}
\end{figure}

To draw a complete quantum circuit explicitly, we define the gate shown in Fig.\ref{fig:control} that performs $U$ operation on the target qubit when the control qubit is $\ket{i}$, otherwise the target qubit remains unchanged. Using this gate, we present a quantum circuit Fig.\ref{fig:two_persons} by which Alice and Bob exchange her $\ket{0}$ and his $\ket{1}$. 

We first consider the case where quantum strategies are not entangled ($\mathcal{J}=\mathcal{J}^\dagger=I$). Generic states which Alice and Bob have can be written as   $\Ket{\psi_A}=A_0\Ket{0}+A_1\Ket{1},\Ket{\psi_B}=B_0\Ket{0}+B_1\Ket{1}$. If a game starts with a general entangled state $\ket{\psi}=\mathcal{K}\ket{\psi_A\psi_B}\in\mathcal{H}_A\otimes\mathcal{H}_B$ between Alice and Bob
\begin{equation}
\label{eq:entangled_state}
    \ket{\psi}=\sum_{i,j\in\{0,1\}}K_{ij}\ket{ij},
\end{equation}
the initial state $\Ket{\psi}\ket{0}_A\ket{0}_B$ eventually becomes 
\begin{equation}
    K_{01}a_1b_1\ket{10}\ket{1}_A\ket{1}_B+\cdots,
\end{equation}
where terms which do not contribute to trade are omitted. So the success probability of the trade is $|K_{01}a_1b_1|^2$. 

Now we consider a trade among the three persons. In a classical system, they have an indivisible quantum commodity called goods 1,2 or 3. Two of the three meet and decide whether to exchange their commodities. If both two persons want to trade, they exchange their commodities. We write each $\Ket{i}$ in the binary form 
\begin{align}
\Ket{1}=\Ket{01}, \Ket{2}=\Ket{10}, \Ket{3}=\Ket{11}.
\end{align}

\begin{figure}[H]
\centering
 \includegraphics[width=\hsize, bb=0 0 824 759]{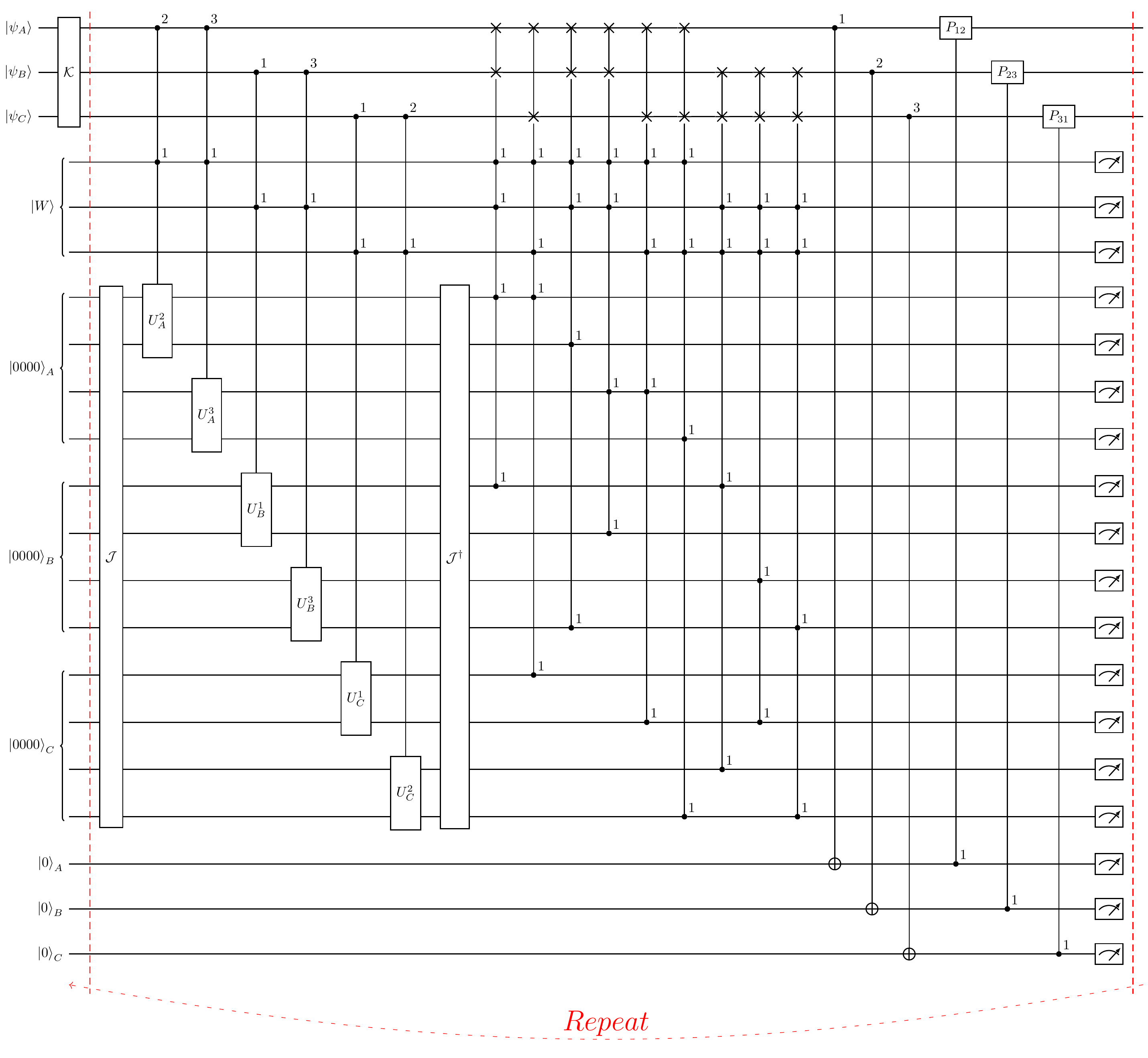}
\caption{Quantum circuit for exchanging quantum states among three persons. Alice, Bob and Charlie exchange $\ket{1},\ket{2}$ and $\ket{3}$, when they have. $\mathcal{K}$ creates entanglement among states $\ket{\psi_A}$, $\ket{\psi_B}$ and $\ket{\psi_C}$. $\mathcal{J}$ yields entanglement among quantum strategies $U_A$, $U_B$ and $U_C$. $P_{ij}$ is a gate for consuming $\ket{i}$ and producing $\ket{j}$. }
    \label{fig:three_persons}
\end{figure}

Suppose each of the three, Alice, Bob and Charlie has a quantum commodity $\Ket{\psi_i}=i_1\Ket{1}+i_2\Ket{2}+i_3\Ket{3} (i=A,B,C)$. Two of them meet and trade pure good states, as in the classical system. We assume a person meets another with the equal probability 1/3 and introduce $\Ket{W}=\frac{1}{\sqrt{3}}(\Ket{110}+\Ket{101}+\Ket{011})$ to describe a person who someone meets. Here $\Ket{110}$ means that Alice meets Bob, $\Ket{101}$ says that Alice meets Charlie, and $\Ket{011}$ indicates that Bob meets Charlie. We illustrate the quantum circuit for a three-person trade in Fig. \ref{fig:three_persons}. The introduction of $\ket{W}$ is one of the most significant differences from the two-person trade diagram (Fig.\ref{fig:two_persons}). For example, Alice's strategy is a sum of tensor products $U_A=\ket{1}\bra{1}\otimes I^{\otimes 4}+\Ket{2}\bra{2}\otimes U^2_A\otimes I^{\otimes 2}+ \Ket{3}\bra{3}\otimes I^{\otimes 2} \otimes U^3_A$, where $U^i_A$ is her strategy for trading a good $i$ when she has $\ket{i}$.

\subsection{\label{sec:entangled_goods}Entangled Quantum Goods, Strategies and Quantum Money as a Medium of Exchange}
\begin{figure}[H]
    \centering
    \includegraphics[scale=0.8,bb=0 0 200 230]{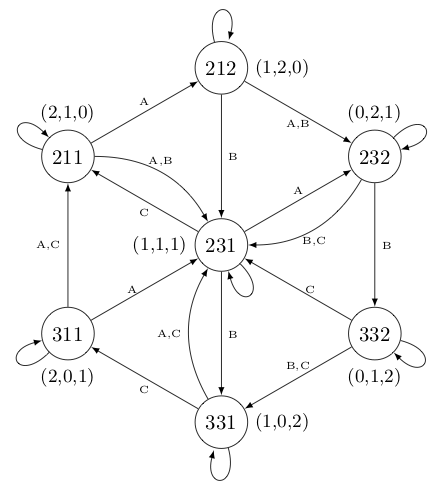}
    \caption{A sequence of numbers in a circle means a state: 231 means Alice, Bob, and Charlie have $\ket{2}$,$\ket{3}$ and $\Ket{1}$. A 3-tuple of numbers outside a circle indicates that of the number of persons who have the consuming quantum commodities. When 211 is shown in a circle, (2,1,0) expresses that two persons have Alice's consuming commodity $\Ket{1}$, only Alice has Bob's consuming good 2, and no one has the state Charlie desires. $231\xrightarrow{C} 211$ means that it is Charlie who gains a reward by an exchange. From the state $\Ket{231}$ to $\Ket{211}$, Charlie receives $\ket{3}$ from Bob, consumes it and produces $\Ket{1}$. }
    \label{fig:diagram}
\end{figure}

We first address the case where quantum goods are entangled  ($\mathcal{K}\neq I^{\otimes3}$) and strategies are not entangled ($\mathcal{J}=I^{\otimes 12}$). Fig. \ref{fig:diagram} presents the diagram of exchange. 

It is convenient to consider a density matrix $\rho$ that describes entangled goods, instead of $\ket{\psi_A\psi_B\psi_C}$. With respect to a given initial state $\ket{\psi_A\psi_B\psi_C}$, we define $\rho$ by
\begin{align}
    \rho&=\mathcal{K}\ket{\psi_A\psi_B\psi_C}\bra{\psi_A\psi_B\psi_C} \mathcal{K}^\dagger\in End(\mathcal{H}_\text{Commodity})\\
    &=\sum_{abc,ijk}C_{abc,ijk}\Ket{abc}\bra{ijk},
\end{align}
where $C_{abc,ijk}$ is a coefficient. In order to discuss an equilibrium of their business, we define an operator acting on the space of quantum commodities  $T:\text{End}(\mathcal{H}_\text{Commodity})\to \text{End}(\mathcal{H}_\text{Commodity})$
\begin{align}
\label{eq:T}
    T(\rho)&=\text{Tr}_{Ancilla} E\left(\rho\otimes\ket{\text{Ancilla}}\bra{\text{Ancilla}}\right) E^\dagger \\
    \Ket{\text{Ancilla}}&=\Ket{W}\Ket{0000}_A\Ket{0000}_B\Ket{0000}_C\Ket{0}_A\Ket{0}_B\Ket{0}_C
\end{align}
where $E:\mathcal{H}_\text{Commodity}\otimes \mathcal{H}_\text{Ancilla} \to \mathcal{H}_\text{Commodity}\otimes \mathcal{H}_\text{Ancilla} $ represents all single-round quantum gate operations in Fig.\ref{fig:three_persons}. We trace out ancilla qubits after all operations are made in a single round, in order to see a transition of quantum commodities. 

We can make a mixed commodity state from a pure commodity state by the quantum circuit shown in Fig. \ref{fig:three_persons}. Using $T$ in \eqref{eq:T}, then we find
\begin{align}
\begin{aligned}
    \text{Tr} T(\rho)=&\text{Tr}_{Commodity} \left( \text{Tr}_{Ancila}   E\left(\rho\otimes\ket{\text{Ancilla}}\bra{\text{Ancilla}}\right)E^\dagger  \right) \\
    =&\text{Tr} E\left(\rho\otimes\ket{\text{Ancilla}}\bra{\text{Ancilla}}\right)E^\dagger=1.
\end{aligned}
\end{align}
Here it is important to note that $T(\rho)$ is not always a density matrix of a pure state. Since we cannot multiply $E=E_1\otimes E_2$ using an unitary operator $E_1$ on $\mathcal{H}_\text{Commodity}$ and $E_2$ on $\mathcal{H}_\text{Ancila}$, we obtain
\begin{align}
\begin{aligned}
    \text{Tr} T(\rho)^2=& \text{Tr}_{Commodity} \left( \text{Tr}_{Ancila}   E\left(\rho\otimes\ket{\text{Ancilla}}\bra{\text{Ancilla}}\right)E^\dagger  \right)^2 \\
    \neq&  \text{Tr}_{Commodity} \text{Tr}_{Ancila}   \left(E\left(\rho\otimes\ket{\text{Ancilla}}\bra{\text{Ancilla}}\right)E^\dagger \right)^2=1.
\end{aligned}
\end{align}
Therefore, in general, $T(\rho)$ is a mixed state because $\text{Tr} T(\rho)^2<1$.

\begin{defn}
We call $\rho$ a steady state if it obeys 
\begin{equation}
    T\rho=\rho. 
\end{equation}
\end{defn}

If a steady state $\rho$ is used for an initial state, the total profit $\pi_i$ \eqref{eq:profit} converges into 
\begin{equation}
\label{eq:expected_payoff}
    \lim_{t\to \infty}\pi_i(t)=\frac{V_i(\rho)}{1-\delta},
\end{equation}
where $V_i$ is a certain function of $\rho$. To give a concrete detail, we first derive Alice's $V_A$. Let $\rho$ be a quantum state computed on the basis $\{\ket{211}, \ket{212}, \ket{231}, \ket{232}, \ket{311},\ket{312}, \ket{331},\ket{332}\}$\footnote{Here we identify $\ket{\psi}\bra{\psi}$ with $\Ket{\psi}$.}. Let $v_i$ be the storage cost that $i$ should pay and $w_i$ be the possibility that $i$ receive $i$. Each $v_i$ can be expressed as 
\begin{align}
    v_A&=(c_2,c_2,c_2,c_2,c_3,c_3,c_3,c_3) &
        v_B&=(c_1,c_1,c_3,c_3,c_1,c_1,c_3,c_3)&
  v_C&=(c_1,c_2,c_1,c_2,c_1,c_2,c_1,c_2)
\end{align}
and $w_i$ depends on $\mathcal{J}$ and $U_i^j$. Using $v_i$ and $w_i$, we find $f_i(t)$\eqref{eq:profit} becomes 
\begin{align}
\begin{aligned}
    f_i(t)&=w_i \cdot T^{t-1} (\rho) \\
    \sum_j c_{ij}p_{ij}(t)&=v_i\cdot T^t  (\rho),
\end{aligned}
\end{align}
where we denote by $\cdot$ the inner product of vectors. We may simply put $T^{-1}(\rho)=0,T^0(\rho)=\rho$. Then the total profit of a player $i$ is 
\begin{align}
\begin{aligned}
\label{eq:profit_A}
\lim_{t\to \infty}\pi_i(t)& =\sum_{\tau=0}^\infty \left(uw_i \cdot T^{\tau-1} (\rho) -v_\cdot T^\tau  (\rho) \right)\delta^{\tau} \\
&=\sum_{\tau=0}^\infty(-v_i  + \delta u w_i)\cdot T^\tau (\rho)\delta^\tau .
\end{aligned}
\end{align}
If $\rho$ is a steady state $T\rho =\rho$, we find $V_i=(-v_i+\delta u w_i)\cdot \rho$ from the equation
\begin{align}
\begin{aligned}
\label{eq:V}
\lim_{t\to \infty}\pi_i(t)&=\sum_{\tau=0}^\infty(-v_i  + \delta u w_i)\cdot \rho\delta^\tau \\
&=\frac{(-v_i+\delta u w_i)\cdot \rho}{1-\delta}
\end{aligned}
\end{align}
Note that $V_i$ dose not explicitly depend on $v_j$ and $w_j$ $(j\neq i)$, but it is related with them via $\rho$. 

\if{Similarly, using the steady state, we can write the total expected profits $V_B,V_C$ of Bob and Charlie as
\begin{align}
\begin{aligned}
V_B&=(-v_B+\delta u w_B)\cdot \rho \\
V_C&=(-v_C+\delta u w_C)\cdot \rho
\end{aligned}
\end{align}
}\fi

\if{
where 
\begin{align}
\begin{aligned}
    v_B&=(c_1,c_1,c_3,c_3,c_1,c_1,c_3,c_3)&
    w_B&=\frac{1}{3}(1,1+s_C,1-s_A,2-s_A,0,s_C,0,1) \\
  v_C&=(c_1,c_2,c_1,c_2,c_1,c_2,c_1,c_2)&
    w_C&=\frac{1}{3}(0,0,s_B,1,1,s_A,1+s_B,1+s_A)   
\end{aligned}
\end{align}
}\fi

\if{
\begin{prop}
$U$ is reducible.
\end{prop}
\begin{prf}
$(U^k)_{6i}=0$ for any $k=1,2,\cdots$ and $i\neq 6$.\Qed{}
\end{prf}
}\fi

Before we describe a generic equilibrium and the flow of entangled quantum goods, let us start with an example which includes a review on the original Kiyotaki-Wright model \cite{10.2307/1832197,Kehoe93}. 
More general case is given in Sec.\ref{sec:theorem}, where Theorem \ref{thm:1} summarizes our main result here. In what follows we assume Alice,Bob and Charlie prefer the following strategies \eqref{eq:ex} 
\begin{align}
\begin{aligned}
\label{eq:ex}
U_A^2&=Y\otimes \left(\sqrt{s_A}I+i\sqrt{1-s_A}Y\right)
&U_A^3&=Y\otimes\left(\sqrt{1-s_A}I+i\sqrt{s_A}Y\right)  \\
U_B^1&=Y\otimes \left( \sqrt{s_B}I+i\sqrt{1-s_B}Y\right)
&U_B^3&=\left( \sqrt{1-s_B}I+i\sqrt{s_B} Y\right)\otimes Y\\
U_C^1&=\left(\sqrt{s_C}I+i\sqrt{1-s_C}Y\right)\otimes Y
&U_C^2&=\left(\sqrt{1-s_C}I+i\sqrt{s_C}Y\right)\otimes Y 
\end{aligned}
\end{align}
Indeed the classical results \cite{10.2307/1832197,Kehoe93} are reproduced when parameters $s_i~(i=A,B,C)$ are chosen appropriately. As the quantum game must reproduce the classical results, we generally require
\begin{equation}
    [\mathcal{J}, U^i_j]=0
\end{equation}
when the strategies are restricted to classical strategies. For this purpose, we use $\mathcal{J}=\exp\left[i\theta Y^{\otimes 12}\right]$ throughout this article. It will be yet another interesting question to study our model when a different $\mathcal{J}$ is used. The strategy is called fundamental when $s_i$ is limited to 1 and speculative when $s_i=0$ \cite{Kehoe93}. Then we identify the concrete form of $w_i$ as follows 
\begin{align}
\begin{aligned}
    w_A&=\frac{1}{3}(2-s_C,1,1-s_C,0,2-s_B,1-s_B,1,0) \\
        w_B&=\frac{1}{3}(1,1+s_C,1-s_A,2-s_A,0,s_C,0,1) \\
    w_C&=\frac{1}{3}(0,0,s_B,1,1,s_A,1+s_B,1+s_A).
\end{aligned}
\end{align}
For example, the first component of $w_A$ can be derived from $\Ket{211}$ as follows. Alice and Bob exchange $\Ket{2}\leftrightarrow\Ket{1}$ with the probability 1 when they meet and, Alice and Charlie exchange $\Ket{2}\leftrightarrow\Ket{1}$ with the probability $1-s_C$ when they meet. Therefore Alice obtains $\Ket{1}$ with the probability $\frac{2-s_C}{3}$ on $\ket{211}$.

One can find that there is a unique steady state for each of $s_A=s_B=s_C=1$ (the fundamental strategy) and $s_A=0, s_B=s_C=1$ (the speculative strategy). Those results are perfectly consistent with the Kiyotaki-Write case \cite{10.2307/1832197}. The steady state at $s_A=s_B=s_C=1$ is
\begin{align}
    \rho = \frac{1}{2} \Ket{211} \bra{211}+\frac{1}{2} \Ket{231} \bra{231},
\end{align}
and the steady state\footnote{In \cite{10.2307/1832197}, the steady state of speculative equilibrium is characterized by   $(p_{12},p_{23},p_{31})=\left(\frac{\sqrt{2}}{2},\sqrt{2}-1,1\right)=(0.707\cdots,0.414\cdots,1)$. In our setup, they are $(p_{12},p_{23},p_{31})=\left(\frac{5}{7},\frac{3}{7},1\right)=(0.714,0.428,1)$.}  at $s_A=0, s_B=s_C=1$ is
\begin{align}
    \rho=\frac{3}{7} \Ket{211} \bra{211}+\frac{2}{7} \Ket{231} \bra{231}+\frac{1}{7} \Ket{311} \bra{311}
    +\frac{1}{7}\Ket{331}\bra{331}.
\end{align}

Moreover there is additional equilibrium at $s_A=s_B=1,s_C=0$, in which $\Ket{1}$ and $\Ket{2}$ can be media of exchange. The corresponding steady state is 
\begin{align}
\label{eq:new_equi}
        \rho=\frac{1}{7}\Ket{211}\Bra{211}+\frac{1}{7}\Ket{212}\Bra{212}+\frac{2}{7}\Ket{231}\Bra{231}+\frac{3}{7}\Ket{232}\Bra{232}. 
\end{align}
Those three equilibira are probabilistic mixture of pure states, which are mixed state without entanglement \cite{PhysRevA.40.4277}. The flow of goods are exhibited in Fig. \ref{fig:lena}. 

\begin{figure}[H]
    \centering
    \includegraphics[width=\hsize,bb=0 0 1199 268]{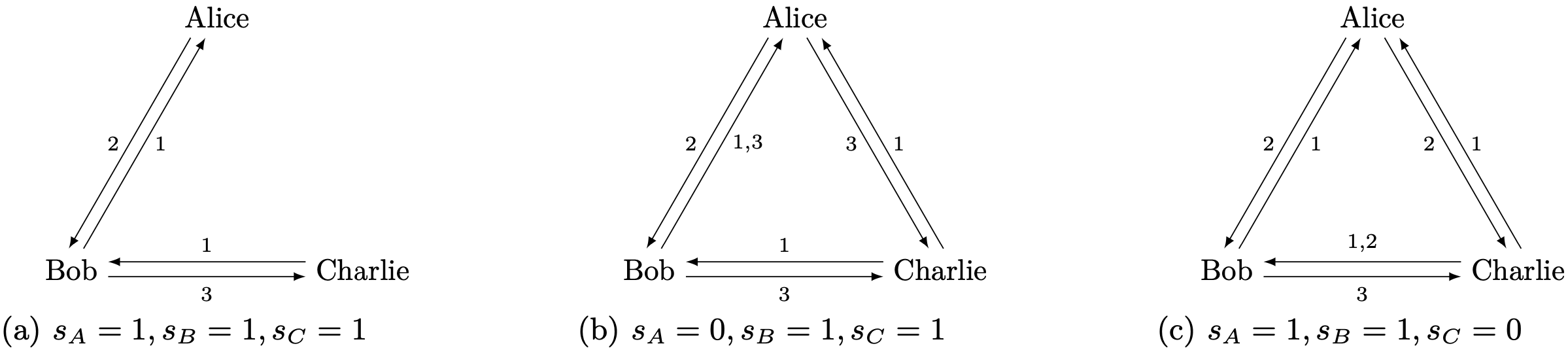}
    \caption{The flow of goods. (a) $\Ket{1}$ is a medium of exchange in the fundamental equilibrium. (b) $\Ket{1}$ and $\Ket{3}$ are media of exchange in the speculative equilibrium. (c) $\Ket{1}$ and $\Ket{2}$ are media of exchange in the equilibrium of \eqref{eq:new_equi}} 
    \label{fig:lena}
\end{figure}

\if{\begin{figure}[H]
  \begin{center}
    \begin{tabular}{c}
      \begin{minipage}{0.32\hsize}
        \begin{center}
          \includegraphics[width=3cm, bb=0 0 100 100]{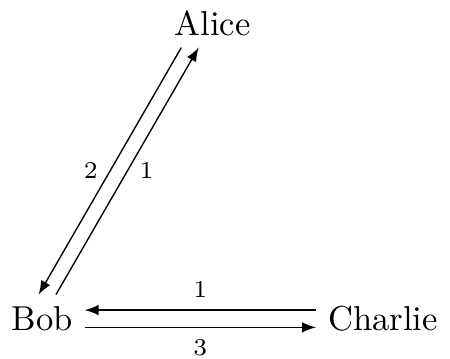}
        \hspace{1.6cm} 
        $\text{(a)}~s_A=1,s_B=1,s_C=1$
        \end{center}
      \end{minipage}

      \begin{minipage}{0.32\hsize}
        \begin{center}
          \includegraphics[width=3cm, bb=0 0 100 100]{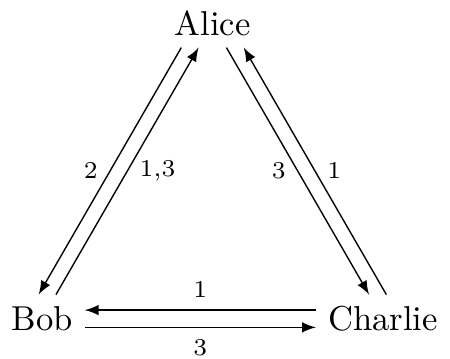}
          \hspace{1.6cm} 
          $\text{(b)}~s_A=0,s_B=1,s_C=1$
        \end{center}
      \end{minipage}

      \begin{minipage}{0.32\hsize}
        \begin{center}
          \includegraphics[width=3cm, bb=0 0 100 100]{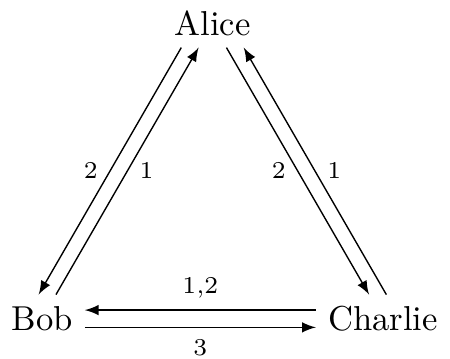}
          \hspace{1.6cm} 
          $\text{(c)}~s_A=1,s_B=1,s_C=0$
        \end{center}
      \end{minipage}

    \end{tabular}
    \caption{The flow of goods. (a) $\Ket{1}$ is a medium of exchange in the fundamental equilibrium. (b) $\Ket{1}$ and $\Ket{3}$ are media of exchange in the speculative equilibrium. (c) $\Ket{1}$ and $\Ket{2}$ are media of exchange in the equilibrium of \eqref{eq:new_equi}} 
    \label{fig:lena}
  \end{center}
\end{figure} }\fi

It is instructive to observe a transition process of an entangled density matrix $\rho$ by $T$. As an example we address $\rho=\Ket{311}\bra{211}$ to help readers follow the computation \eqref{eq:bob_charlie}. By exchange, consumption and production, the state $\Ket{211}$ becomes 
\begin{align}
\begin{aligned}\label{eq:211C}
    &\Ket{211} \Ket{W} \Ket{0000}_A\otimes\Ket{0000}_B\otimes\Ket{0000}_C \otimes\Ket{0}_A\Ket{0}_B\Ket{0}_C\\
    \to &-\Ket{231} \frac{1}{\sqrt{3}}\Ket{110}\Ket{1}_A(\sqrt{s_A}\Ket{0}_A-\sqrt{1-s_A}\Ket{1}_A)\Ket{00}_A\otimes \Ket{1}_B(\sqrt{s_B}\Ket{0}_B-\sqrt{1-s_B}\Ket{1}_B)\Ket{00}_B
    \otimes \Ket{0000}_C \Ket{1}_A\Ket{1}_B\Ket{0}_C \\
    &-\Ket{211}\frac{1}{\sqrt{3}}\Ket{101}\Ket{1}_A(\sqrt{s_A}\Ket{0}_A-\sqrt{1-s_A}\Ket{1}_A)\Ket{00}_A\otimes \Ket{0000}_B \otimes \sqrt{s_C}\Ket{0}_C\Ket{100}_C \Ket{0}_A\Ket{0}_B\Ket{0}_C \\
    &+\Ket{212}\frac{1}{\sqrt{3}}\Ket{101}\Ket{1}_A(\sqrt{s_A}\Ket{0}_A-\sqrt{1-s_A}\Ket{1}_A)\Ket{00}_A \otimes \Ket{0000}_B \otimes \sqrt{1-s_C}\Ket{1}_C\Ket{100}_C \Ket{1}_A\Ket{0}_B\Ket{0}_C \\
    &-\Ket{211}\frac{1}{\sqrt{3}}\Ket{011}\Ket{0000}_A \otimes \Ket{1}_B(\sqrt{s_B}\Ket{0}_B-\sqrt{1-s_B}\Ket{1}_B)\Ket{00}_B\otimes  (\sqrt{s_C}\Ket{0}_C-\sqrt{1-s_C}\Ket{1}_C)\Ket{100}_C \Ket{0}_A\Ket{0}_B\Ket{0}_C
\end{aligned} 
\end{align}

Similarly, the state $\Ket{311}$ is mapped to 
\begin{align}
\begin{aligned}\label{eq:311C}
    &\Ket{311} \Ket{W} \Ket{0000}_A\otimes\Ket{0000}_B\otimes\Ket{0000}_C\Ket{0}_A\Ket{0}_B\Ket{0}_C \\
    \to &-\Ket{311} \frac{1}{\sqrt{3}}\Ket{110}\Ket{001}_A(\sqrt{1-s_A}\Ket{0}_A-\sqrt{s_A}\Ket{1}_A)  \otimes \sqrt{s_B}\Ket{1000}_B
    \otimes \Ket{0000}_C  \Ket{0}_A\Ket{0}_B\Ket{0}_C\\
    &+\Ket{231} \frac{1}{\sqrt{3}}\Ket{110}\Ket{001}_A(\sqrt{1-s_A}\Ket{0}_A-\sqrt{s_A}\Ket{1}_A) \otimes \sqrt{1-s_B}\Ket{1100}_B
    \otimes \Ket{0000}_C  \otimes\Ket{1}_A\Ket{0}_B\Ket{0}_C\\
    &-\Ket{211}\frac{1}{\sqrt{3}}\Ket{101}\Ket{001}_A(\sqrt{1-s_A}\Ket{0}_A-\sqrt{s_A}\Ket{1}_A) \otimes \Ket{0000}_B\otimes(\sqrt{s_C}\Ket{0}-\sqrt{1-s_C}\Ket{1})\Ket{100}_C \Ket{1}_A\Ket{0}_B\Ket{1}_C \\
    &-\Ket{311}\frac{1}{\sqrt{3}}\Ket{011}\Ket{0000}_A \otimes\Ket{1}_B(\sqrt{s_B}\Ket{0}_B-\sqrt{1-s_B}\Ket{1}_B)\Ket{00}_B \otimes  (\sqrt{s_C}\Ket{0}_C-\sqrt{1-s_C}\Ket{1}_C)\Ket{100}_C \Ket{0}_A\Ket{0}_B\Ket{0}_C
\end{aligned}
\end{align}
By tracing out the ancilla qubits, we obtain 
\begin{align}
    T(\Ket{311}\bra{211})=\frac{1}{3} \Ket{311}\bra{211}
\end{align}
and
\begin{equation}
\label{eq:entangled}
  \lim_{t\to\infty}T^t  (\Ket{311}\bra{211})=0. 
\end{equation}
By repeating the same calculation, one can find \eqref{eq:entangled} is true for any entangled state $\ket{abc}\bra{ijk}$ $(\ket{abc}\neq\ket{ijk})$. Therefore it is sufficient to consider the $\ket{abc}=\ket{ijk}$ cases. Then the main components of the transition matrix $T$ can be represented with basis $\{\ket{211}, \ket{212}, \ket{231}, \ket{232}, \ket{311},\ket{312}, \ket{331},\ket{332}\}$ in such a way that
\begin{equation}
T=
    \begin{pmatrix}
    \frac{1+s_C}{3}&0&\frac{s_B}{3}&0&\frac{1}{3}&\frac{s_A}{3}&0&0\\
    \frac{1-s_C}{3}&\frac{2-s_C}{3}&0&0&0&0&0&0\\
    \frac{1}{3}&\frac{s_C}{3}&\frac{s_A+1-s_B+s_C}{3}&\frac{1}{3}&\frac{1-s_B}{3}&0&\frac{1}{3}&\frac{s_A}{3}\\
    0&\frac{1}{3}&\frac{1-s_C}{3}&\frac{s_A+1}{3}&0&\frac{1-s_B}{3}&0&0\\
    0&0&0&0&\frac{1+s_B}{3}&0&\frac{s_B}{3}&0\\
    0&0&0&0&0&\frac{2-s_A+s_B-s_C}{3}&0&0\\
    0&0&\frac{1-s_A}{3}&0&0&\frac{s_C}{3}&\frac{2-s_B}{3}&\frac{1}{3}\\
    0&0&0&\frac{1-s_A}{3}&0&0&0&\frac{2-s_A}{3}
    \end{pmatrix}.
\end{equation}
For example, $\ket{211}$ is mapped into $\ket{212}$ and $\ket{231} $ with the probability $\frac{1-s_C}{3}$ and $\frac{1}{3}$, respectively. $\ket{312}$ is mapped to itself by the $T$ operation, so $T_{6i}= 0$ unless $i=6$. From the Perron-Frobenius theorem, the maximal eigenvalue of the non-negative matrix $T$ is $1$ and the components of the corresponding eigenvector, which may not be uniquely determined, is non-negative.

\begin{figure}[H]
    \begin{center}
    \includegraphics[width=7cm, bb=0 0 460 342]{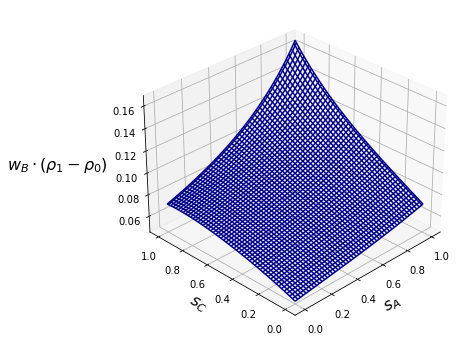}
    \end{center}
    \caption{The $(s_A,s_C)$-dependence of $w_B\cdot (\rho_1-\rho_0)$, which is larger than 0 for any pair of $s_A,s_C\in [0,1]$.}
    \label{fig:Bob}
\end{figure}

Now let us discuss equilibria of quantum strategies. We assume $s_B=1$ because of the inequality $V_B(s_A,0,s_C)<V_B(s_A,1,s_C)$ for all $s_A,s_C \in [0,1]$. Let $\rho_0,\rho_1$ be steady states at $s_B=0,1$, respectively. Bob keeps $\ket{3}$ if $s_B=0$. Therefore we obtain 
\begin{align}
    v_B\cdot \rho_0=c_3 \geq v_B\cdot \rho_1
\end{align}
and 
\begin{align}
    V_B(s_A,1,s_C)-V_B(s_A,0,s_C)&=-v_B\cdot (\rho_1-\rho_0)+\delta u w_B\cdot (\rho_1-\rho_0) \\
    &=(c_3-v_B\cdot \rho_1)+\delta u w_B\cdot (\rho_1-\rho_0). 
\end{align}
$c_3-v_B\cdot \rho\geq 0$ and the second term is grater than 0 for $s_A,s_C\in [0,1]$ as Fig.\ref{fig:Bob} shows. Therefore $V_B(s_A,1,s_C)>V_B(s_A,0,s_C)$ is satisfied. 

\begin{figure}[H]
    \centering
    \includegraphics[width=9cm, bb=0 0 250 320]{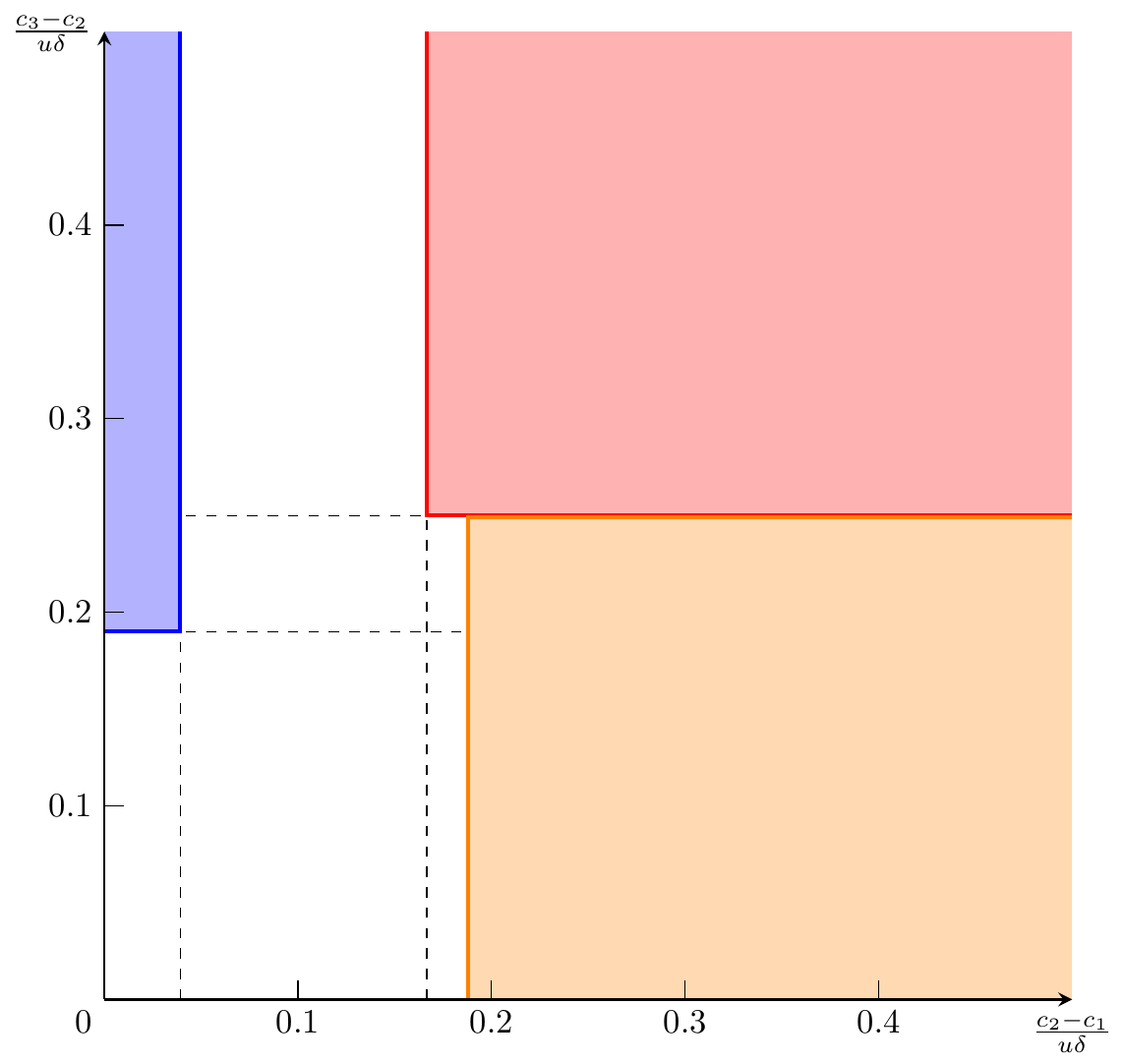}
    \caption{Equilibrium strategies of Alice and Charlie when $s_B=1$.}
    \label{fig:optimal}
\end{figure}

\begin{figure}[H]
\centering
\includegraphics[width=\hsize,bb= 0 0 1349 374]{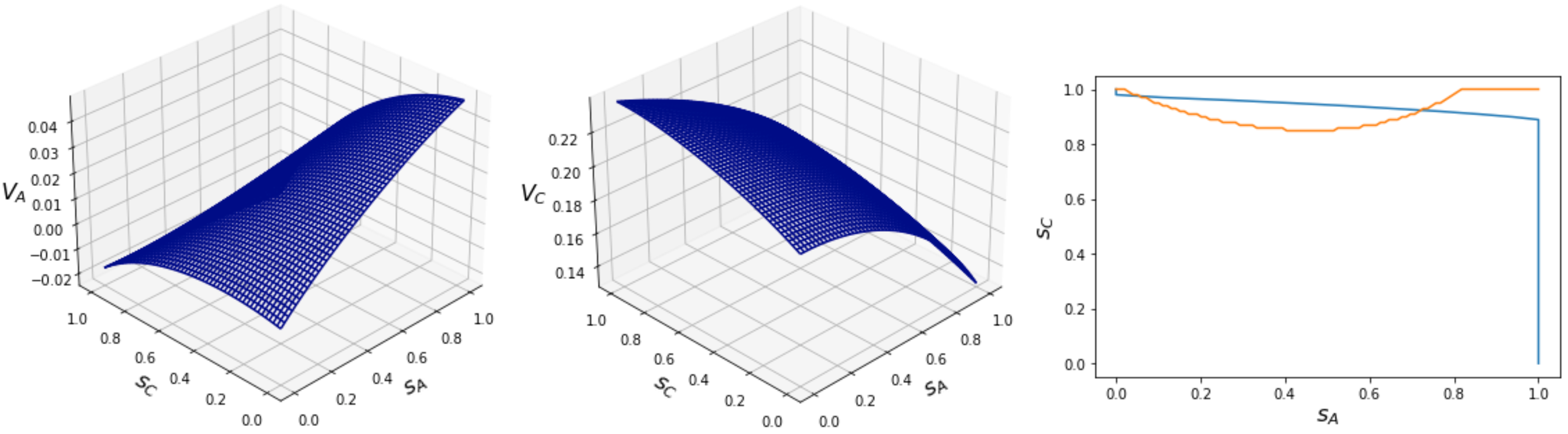}
\if{\begin{minipage}{0.32\hsize}
    \centering
     \includegraphics[width=\hsize,bb=0 0 363 319]{Figure1_1.png}
\end{minipage}
\begin{minipage}{0.32\hsize}
    \centering
     \includegraphics[width=\hsize, bb=0 0 363 319]{Figure2_1.png}
\end{minipage}
\begin{minipage}{0.32\hsize}
    \centering
    \includegraphics[width=\hsize,bb=0 0 363 319]{Figure3_1.png}
\end{minipage} }\fi
\caption{Equilibrium strategies of Alice and Charlie at a point $\frac{c_2-c_1}{u\delta}=0.19,\frac{c_3-c_2}{u\delta}=0.23$ near the boundary between the red and the white domains in Fig.\ref{fig:optimal}. The blue line in the right figure corresponds to the optimal $s_A$ against $s_C$ and the red line indicates the optimal $s_C$ against $s_A$. There are three crossing points. Charlie desires his best situation $(s_A,s_C)=(0,1)$, which is not best for Alice. The right figure also suggests that $s_A=1$ is Alice's optimal strategy if $s_C\in[0,0.9].$}
\label{fig:three}
\end{figure}

An optimal pair of $s_A,s_C$ depends on $c_2-c_1, c_3-c_2$ and $\delta u$. Equilibrium strategies of Alice and Charlie are shown in Fig.\ref{fig:optimal}. $(s_A,s_C)=(1,1), (0,1), (1,0)$ are the equilibria in the red, orange and blue domains, respectively. In the white region, at least one of Alice and Charlie uses a mixed strategy $s_i\neq 0,1$ for an equilibrium. Note that there might be multiple equilibria near a boundary. For example, as illustrated in Fig.\ref{fig:three}, there are three equilibria at a point $\frac{c_2-c_1}{u\delta}=0.19,\frac{c_3-c_2}{u\delta}=0.23$ near the boundary between orange and white.

\subsection{\label{sec:theorem}Emergence of Commodity Money from Quantum Exchange}
In this section we generalized our discussion in Sec.\ref{sec:entangled_goods}. In addition to entangled quantum goods, we also consider quantum strategies that do not always commute $\mathcal{J}=\exp\left[i\theta Y^{\otimes 12}\right]$, hence strategies of Alice, Bob and Charlie are generally entangled. For strategies in Fig.\ref{fig:three_persons}, we impose $\bra{00}U_i^j\Ket{00}=\bra{11}U_i^j\Ket{11}=0~(i=\text{A,B,C},j\in\{1,2,3\})$ in order to secure the independence of strategies for different goods. The strategies consist of
\begin{align}
\begin{aligned}
\bar{U}_A^2&=U_A^2\otimes I^{\otimes 2},~\bar{U}_A^3=I^{\otimes 2} \otimes U_A^3\\
\bar{U}_B^1&=U_B^1\otimes I^{\otimes 2},~\bar{U}_B^3=I^{\otimes 2} \otimes U_B^3\\
\bar{U}_C^1&=U_C^1\otimes I^{\otimes 2},~\bar{U}_C^2=I^{\otimes 2} \otimes U_C^2
\end{aligned}
\end{align}
We first consider $\mathcal{J}= I^{\otimes 12}$. With respect to a state $\Ket{abc}$ of quantum goods, the strategies act on the ancillary term as
\begin{align}
\begin{aligned}
\Ket{W}\Ket{0000}_A\Ket{0000}_B\Ket{0000}_C
\to& \frac{1}{\sqrt{3}}\Ket{110}(\bar{U}_A^a\otimes \bar{U}_B^b\otimes I^{\otimes 4})\Ket{0000}_A\Ket{0000}_B\Ket{0000}_C \\
&+ \frac{1}{\sqrt{3}}\Ket{101}(\bar{U}_A^a\otimes I^{\otimes 4}\otimes \bar{U}_C^c)\Ket{0000}_A\Ket{0000}_B\Ket{0000}_C \\
&+ \frac{1}{\sqrt{3}}\Ket{011}(I^{\otimes 4}\otimes \bar{U}_B^b \otimes \bar{U}_C^c)\Ket{0000}_A\Ket{0000}_B\Ket{0000}_C 
\end{aligned}
\end{align}

By definition, each player has 4 qubits. If two persons meet, at least one of the first half two qubits or the second half two qubits becomes $\Ket{1}$. If a person meets no one, all of his/her 4 qubits are $\Ket{0}$. For example, if Alice $\ket{2}$ meet someone, one of her first two qubits is $\ket{1}$. Similarly if she has $\Ket{3}$ and meet someone, one of her second half two qubits becomes $\Ket{1}$. Therefore  $T(\Ket{abc}\bra{ijk})\neq0$ obeys if and only if one of the following conditions satisfies 
\begin{align}
\begin{aligned}
\label{eq:conditions}
\Ket{b}=\Ket{j},\Ket{c}=\Ket{k} \\
\Ket{a}=\Ket{i},\Ket{c}=\Ket{k} \\
\Ket{a}=\Ket{i},\Ket{b}=\Ket{j}
\end{aligned}
\end{align}
In addition, if only one player has a different good, say $\Ket{a}\neq \Ket{i},\Ket{b}=\Ket{j},\Ket{c}=\Ket{k}$, non-zero terms are given when Bob and Charlie meet 
\begin{align}
\label{eq:bob_charlie}
T(\ket{abc}\bra{ibc})=\frac{\alpha}{3}\ket{a\contraction{}{b}{}{c}
bc}\bra{i\contraction{}{b}{}{c}
bc}+\frac{1-\alpha}{3}\Ket{abc}\bra{ibc},
\end{align}
where $\ket{a\contraction{}{b}{}{c}
bc}$ means the state after the exchange $\Ket{b}\leftrightarrow \ket{c}$ and $\alpha\in [0,1]$ is the success probability of the exchange. As a general result, any entangled term becomes smaller with the factor $\frac{1}{3}$, hence  $\lim_{t\to\infty}T^t(\ket{abc}\bra{ibc})=0$ holds. 

Now let us consider the case $\mathcal{J}=\exp{\left[i \theta Y^{\otimes 12}\right]}$. The ancillary term is mapped to 
\begin{align}
\begin{aligned}
\Ket{W}\Ket{0000}_A\Ket{0000}_B\Ket{0000}_C
\to& \frac{1}{\sqrt{3}}\Ket{110}\mathcal{J}^\dagger (\bar{U}_A^a\otimes \bar{U}_B^b\otimes I^{\otimes 4})\mathcal{J}\Ket{0000}_A\Ket{0000}_B\Ket{0000}_C \\
&+ \frac{1}{\sqrt{3}}\Ket{101}\mathcal{J}^\dagger(\bar{U}_A^a\otimes I^{\otimes 4}\otimes \bar{U}_C^c)\mathcal{J}\Ket{0000}_A\Ket{0000}_B\Ket{0000}_C \\
&+ \frac{1}{\sqrt{3}}\Ket{011}\mathcal{J}^\dagger(I^{\otimes 4}\otimes \bar{U}_B^b \otimes \bar{U}_C^c)\mathcal{J}\Ket{0000}_A\Ket{0000}_B\Ket{0000}_C 
\end{aligned}
\end{align}
The terms sandwiched by $\mathcal{J}^\dagger,\mathcal{J}$ can be decomposed into a sum of terms which commute or anti-commute with $\mathcal{J}$
\begin{align}
    \mathcal{J}^\dagger((\text{commutative})+(\text{anti-commutative}) )\mathcal{J}
    =(\text{commutative})-(\text{anti-commutative})\mathcal{J}^2. 
\end{align}
The "(commutative)" part produces the same states of $\mathcal{J}=I^{\otimes 12}$ and the "(anti-commutative)" part gives different states which consist of the following terms: if a player meets someone, at least one of the first or second half two qubits is $\ket{0}$ and, if a player dose not meet anyone, all of his/her 4 qubits are $\Ket{1}$. Therefore $T(\Ket{abc}\Bra{ijk})\neq 0$ if and only if one of the conditions \eqref{eq:conditions} is satisfied. Transition of states before tracing out the corresponding ancilla qubits is summarized in Table. \ref{tab:transition}. 

\begin{table}[H]
    \centering
    \begin{tabular}{c|c|c|c|}\small{
{Goods}$\backslash$Encounter}
 & $\ket{110}$ & $\ket{101}$ & $\ket{011}$ \\ \hline
$\ket{211}$ & \begin{tabular}{c} $\ket{\star\star00}_A\ket{\star\star00}_B\ket{0000}_C $ \\ $\ket{\ast\ast11}_A\ket{\ast\ast11}_B\ket{1111}_C $\end{tabular} & \begin{tabular}{c} $\ket{\star\star00}_A\ket{0000}_B\ket{\star\star00}_C  $\\$ \ket{\ast\ast11}_A\ket{1111}_B\ket{\ast\ast11}_C$ \end{tabular} & \begin{tabular}{c}$ \ket{0000}_A\ket{\star\star00}_B\ket{00\star\star}_C  $\\$ \ket{1111}_A\ket{\ast\ast11}_B\ket{11\ast\ast}_C$ \end{tabular} \\ \hline
$\ket{212}$ & \begin{tabular}{c} $\ket{\star\star00}_A\ket{\star\star00}_B\ket{0000}_C $ \\$ \ket{\ast\ast11}_A\ket{\ast\ast11}_B\ket{1111}_C$ \end{tabular} & \begin{tabular}{c} $\ket{\star\star00}_A\ket{0000}_B\ket{00\star\star}_C$  \\$ \ket{\ast\ast11}_A\ket{1111}_B\ket{11\ast\ast}_C $\end{tabular} & \begin{tabular}{c}$ \ket{0000}_A\ket{\star\star00}_B\ket{\star\star00}_C  $\\ $\ket{1111}_A\ket{\ast\ast11}_B\ket{\ast\ast11}_C $\end{tabular} \\ \hline
$\ket{231}$ & \begin{tabular}{c} $\ket{\star\star00}_A\ket{00\star\star}_B\ket{0000}_C $ \\ $\ket{\ast\ast11}_A\ket{11\ast\ast}_B\ket{1111}_C $\end{tabular} & \begin{tabular}{c}$ \ket{\star\star00}_A\ket{0000}_B\ket{\star\star00}_C $ \\$ \ket{\ast\ast11}_A\ket{1111}_B\ket{\ast\ast11}_C$ \end{tabular} & \begin{tabular}{c}$ \ket{0000}_A\ket{00\star\star}_B\ket{00\star\star}_C  $\\$ \ket{1111}_A\ket{11\ast\ast}_B\ket{11\ast\ast}_C $\end{tabular} \\ \hline
$\ket{232}$ & \begin{tabular}{c} $\ket{\star\star00}_A\ket{00\star\star}_B\ket{0000}_C  $\\ $\ket{\ast\ast11}_A\ket{11\ast\ast}_B\ket{1111}_C$ \end{tabular} & \begin{tabular}{c} $\ket{\star\star00}_A\ket{0000}_B\ket{00\star\star}_C  $\\ $\ket{\ast\ast11}_A\ket{1111}_B\ket{11\ast\ast}_C$\end{tabular} & \begin{tabular}{c} $\ket{0000}_A\ket{00\star\star}_B\ket{\star\star00}_C  $\\ $\ket{1111}_A\ket{11\ast\ast}_B\ket{\ast\ast11}_C $\end{tabular} \\ \hline
$\ket{311}$ & \begin{tabular}{c}$ \ket{00\star\star}_A\ket{\star\star00}_B\ket{0000}_C$  \\ $\ket{11\ast\ast}_A\ket{\ast\ast11}_B\ket{1111}_C $\end{tabular} & \begin{tabular}{c} $\ket{00\star\star}_A\ket{0000}_B\ket{\star\star00}_C  $\\$ \ket{11\ast\ast}_A\ket{1111}_B\ket{\ast\ast11}_C$ \end{tabular} & \begin{tabular}{c} $\ket{0000}_A\ket{\star\star00}_B\ket{00\star\star}_C  $\\ $\ket{1111}_A\ket{\ast\ast11}_B\ket{11\ast\ast}_C $\end{tabular} \\ \hline
$\ket{312}$ & \begin{tabular}{c} $\ket{00\star\star}_A\ket{\star\star00}_B\ket{0000}_C  $\\$ \ket{11\ast\ast}_A\ket{\ast\ast11}_B\ket{1111}_C $\end{tabular} & \begin{tabular}{c} $\ket{00\star\star}_A\ket{0000}_B\ket{00\star\star}_C  $\\ $\ket{11\ast\ast}_A\ket{1111}_B\ket{11\ast\ast}_C $\end{tabular} & \begin{tabular}{c}$ \ket{0000}_A\ket{\star\star00}_B\ket{\star\star00}_C $ \\ $\ket{1111}_A\ket{\ast\ast11}_B\ket{\ast\ast11}_C $\end{tabular} \\ \hline
$\ket{331}$ & \begin{tabular}{c} $\ket{00\star\star}_A\ket{00\star\star}_B\ket{0000}_C $\\$ \ket{11\ast\ast}_A\ket{11\ast\ast}_B\ket{1111}_C$ \end{tabular} & \begin{tabular}{c} $\ket{00\star\star}_A\ket{0000}_B\ket{\star\star00}_C  $\\ $\ket{11\ast\ast}_A\ket{1111}_B\ket{\ast\ast11}_C$ \end{tabular} & \begin{tabular}{c}$ \ket{0000}_A\ket{00\star\star}_B\ket{00\star\star}_C $ \\ $\ket{1111}_A\ket{11\ast\ast}_B\ket{11\ast\ast}_C $\end{tabular} \\ \hline
$\ket{332}$ & \begin{tabular}{c} $\ket{00\star\star}_A\ket{00\star\star}_B\ket{0000}_C  $\\$ \ket{11\ast\ast}_A\ket{11\ast\ast}_B\ket{1111}_C $\end{tabular} & \begin{tabular}{c} $\ket{00\star\star}_A\ket{0000}_B\ket{00\star\star}_C  $\\ $\ket{11\ast\ast}_A\ket{1111}_B\ket{11\ast\ast}_C$\end{tabular} & \begin{tabular}{c}$ \ket{0000}_A\ket{00\star\star}_B\ket{\star\star00}_C $ \\ $\ket{1111}_A\ket{11\ast\ast}_B\ket{\ast\ast11}_C $\end{tabular} \\ \hline
    \end{tabular}
    \caption{Correspondence between quantum goods and the terms associated with $\ket{110}, \ket{101}$ and $\ket{011}$ after exchange, consumption and creation. Terms contained in (commutative) are shown in the upper lows and terms contained in (anti-commutative) are shown in the lower and upper lows. $\star\star$ means at least one of the two qubits is 1 and $\ast \ast$ means at least one of the two qubits is 0. For example, looking at \eqref{eq:211C} we find that $\Ket{211}$ is mapped to $
        \Ket{1000}_A\ket{1000}_B\Ket{0000}_C,\ \Ket{1000}_A\Ket{1100}\Ket{0000}_C, \       \Ket{1100}_A\ket{1000}_B\Ket{0000}_C, \ \Ket{1100}_A\Ket{1100}\Ket{0000}_C
    $ after all possible exchange, consumption and creation of goods that may happen when Alice and Bob meet $(\ket{110})$. This situation is simply expressed as $\ket{\star\star00}_A\ket{\star\star00}_B\ket{0000}_C $. Moreover terms in a lower low appear if $\theta \neq 0$. See \eqref{eq:311Q} for example.}
    \label{tab:transition}
\end{table}

We give a more general argument which assures quantum states can be a medium of exchange. We find entangled stats and classical states have a different durability. Our main goal in this section is to show the following statement.  
 \begin{thm}
\label{thm:1}
Suppose $\mathcal{J}=\exp\left[i\theta Y^{\otimes 12}\right]$ and $\bra{00}U_i^j\Ket{00}=\bra{11}U_i^j\Ket{11}=0$. For any $\theta\in[0,2\pi]$ and any quantum commodity state $\rho$, there is a classical mixed state $\rho_c$ such that 
\begin{equation}
    \lim_{t\to\infty} T^{t}\rho=\rho_c
\end{equation}. 
\end{thm}

\begin{prf}
We consider transition of a state $\ket{abc}\bra{ijk}$ by $T$. Consulting Table.\ref{tab:transition}, one can verify the following formula one-by-one for any $\ket{abc}\neq \ket{ijk}$
\begin{equation}
T(\ket{abc}\bra{ijk})=
\begin{cases}
\frac{1}{3}\ket{abc}^\star\bra{ijk}^\star&(a=i,b=j,c\neq k), ~(a=i,b\neq j, c=k),~\text{or}~(a\neq i,b=j,c=k)\\
0&\text{otherwise}
\end{cases}
\end{equation}
where $\ket{abc}^*\bra{ijk}^*$ is a certain normalized state. Important fact is that operating $T$ to any entanglement term makes it smaller with the factor $\frac{1}{3}$ at most. Therefore we conclude $\lim_{t\to\infty }T^t(\Ket{abc}\bra{ijk})=0$ for any $\ket{abc}\neq \ket{ijk}$. 
\if{
To prove the statement, we explain that there are no entangled steady states. With respect to a generic state $\Ket{abc}\neq \Ket{ijk}$, one can find 
\begin{align}
    T(\Ket{abc}\bra{ijk})=\begin{cases}
    \frac{1}{3} \Ket{abc}^\star \bra{ijk}^\star & (a= i,b= j, c\neq k \ or\ a= i,b\neq j, c= k \ or\ a\neq i,b= j, c= k) \\
    0 & (otherwise),
  \end{cases}
\end{align}
where $\Ket{abc}^\star\bra{ijk}^\star$ is the normalized state obtained by $T$. 
}\fi
\if{よって$\Ket{abc}\neq \Ket{ijk}$の基底のうち，全て異なるか，1qubitだけ同じ基底は$T$によって０となる．2qubitだけ同じ基底は次の期も2qubitだけ同じ基底で毎期$\frac{1}{3}$がかかるから，$T$を掛け続けることで無視できる．よって$\Ket{abc}=\Ket{ijk}$の基底のみを考えれば十分である．}\fi 

When we restrict ourselves to the cases of $\Ket{abc}=\Ket{ijk}$, then $T(\ket{ijk}\bra{ijk})$ is a non-negative matrix. Note that $T$ is a stochastic matrix. If two persons have the same goods, which corresponds to a vertex state of the hexagon in Fig. \ref{fig:diagram}, the state is mapped to itself with the probability $1/3$. Suppose all three persons have different goods. Based on their strategies, it turns out $\ket{312}$ is mapped to itself with the probability 1, in which case $\ket{312}$ is a fixed point, or it is mapped to another vertex state. If $\ket{312}$ is mapped to another state, then it never returns to $\ket{312}$. Therefore we find it vanishes $\lim_{t\to \infty }T^t (\Ket{312}\Bra{312})= 0$. Regarding $\ket{231}$, it can be mapped to itself and other states can be mapped to $\ket{231}$. Based on strategies, the 7 states or the 6 vertex states in the hexagon in Fig. \ref{fig:diagram} form a closed, ergodic and reducible Markov chain. For such a Markov chain, we can find a family of closed, ergodic and irreducible subchains as follows. Up to similarity transformations, our transition matrix for a given Markov chain can be written as 
\begin{align}\label{eq:reducible matrix}
    T=\left(\begin{array}{ccc|ccc|cccc}
T_{1_1} &  & \ast & \multicolumn{7}{r}{0}  \\
 & \ddots &  & \multicolumn{7}{r}{}   \\
0 &  & T_{1_k}  & \multicolumn{7}{r}{} \\ \cline{1-6}
\multicolumn{3}{l|}{}  &T_{2_1}  &    &  \ast  &   \multicolumn{4}{r}{}   \\
\multicolumn{3}{l|}{}   &     &  \ddots  &    &   \multicolumn{4}{r}{}  \\
\multicolumn{3}{l|}{}   & 0   &    &  T_{2_l}  &   \multicolumn{4}{r}{}  \\ \cline{4-6}
 \multicolumn{6}{l}{}&     \ddots  &   \multicolumn{3}{r}{}   \\ \cline{8-10}
 \multicolumn{7}{l|}{}   & T_{N_1} &  & \ast \\
 \multicolumn{7}{l|}{}   &  & \ddots &  \\
 \multicolumn{7}{l|}{0}  & 0 &  & T_{N_m} 
 \end{array} \right)
\end{align}
where $T_{i_j}$ are transition matrices of an irreducible Markov subchain in a given reducible Markov chain and the block matrices in boxes are transition matrices of closed Markov subchains. In particular, $T_{i_1}~(i=1,\cdots,N)$ are transition matrices of closed irreducible and ergodic Markov subchains. According to the convergence theorem for Markov chains, each $T_{i_1}$ has a stationary distribution $\rho_i$. By repeating the deals, the initial state $\rho$ converges into a sum of the stationary distributions 
\begin{align}
    \rho=&a_1 \rho|_{T_{1_1},\cdots , T_{1_k}}+ a_2\rho|_{T_{2_1},\cdots ,T_{2_l}}+ \cdots +a_N\rho|_{T_{N_1},\cdots ,T_{N_m}}  \\
    \to& \sum_{i=1}^{N}a_i\rho_i =\rho_c 
\end{align}
where $\{a_i\}_{i=1}^N$ is a certain family of non-negative numbers that obey $\sum_{i=1}^N a_i=1$ and $\rho|_{T_{i_1},\cdots,T_{i_n}}$ is a projection state of $\rho$ onto a state transited by the corresponding $i$th block matirx consisting of $T_{i_1},\cdots,T_{i_n}$. 

\if{
The case of \eqref{eq:reducible matrix} is suffices since $T$ is expressed as a matrix where these matrices, such as \eqref{eq:reducible matrix}, are placed diagonally in general. If T is written by the left matrix of \eqref{eq:reducible matrix}, then $T_1$ and $T_2$ each have fixed points $\rho_1$ and $ \rho_2$ by the convergence theorem for Markov chains. If T is written by the right matrix of \eqref{eq:reducible matrix}, then the Markov chain of $T_1$ is attractive. By repeating the deal, the initial state transits from $T_2$ to $T_1$ and converge the fixed point of $T_1$ at final.}\fi 

\if{ Reducible Markov chains vanish by infinite repetition of the operation.\fix{ 可約な部分の遷移行列は例えば
\begin{align}
    T=\left(\begin{array}{cc}
        T_1 & 0 \\
        0 & T_2
    \end{array}\right),\left(
    \begin{array}{cc}
        T_1 & \ast \\
        0 & T_2
    \end{array}\right)
\end{align}
のようにすることができます．一般にはこれらが対角な部分にたくさん配置されます．$T_1,T_2$が既約なマルコフ連鎖の遷移行列です．前者は$T_1,T_2$がそれぞれで閉じているので，マルコフ収束定理によってそれぞれに固定点$\rho_1,\rho_2$があります．後者は$T_1$のMarkov連鎖が吸引的で，取引を繰り返すと$T_2$から$T_1$に遷移して最終的に$T_1$の固定点$\rho_1$に収束します． By applying the convergence theorem
}}\fi

Therefore for any $\rho$, one can find a fixed state $\rho_c$ such that $\lim_{t\to\infty}T^t\rho=\rho_c$. Note that such a fixed point is probabilistic mixture of pure states, thereby classical state. Therefore, in general, repetitions of an exchange of quantum commodities make entangled states gradually disappear and, thereby only a classical mixed state $\rho_c$ survives. This completes the proof of Theorem \ref{thm:1}. 
\qed
\end{prf}

\subsection{\label{sec:entangled_strategy}Emergence of Coalition and Quantum Effects on Deals}
\begin{figure}[H]
    \centering
    \includegraphics[width=8cm,bb=0 0 150 150]{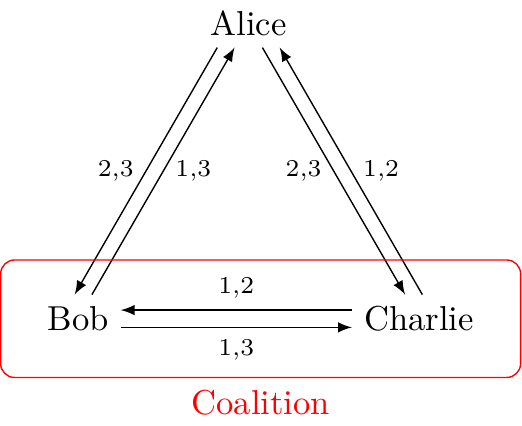}
    \caption{The flow of goods. The difference from the classical case is that 3,2,1 are communicated between Alice and Bob, Bob and Charlie, and Charlie and Alice, respectively. Bob and Charlie can cooperate that does not decrease their profits. This does not happen in a classical situation.}
    \label{fig:flow2}
\end{figure}

Morgenstern and Neumann assert that a three-person decision making should be completely different from a two-person decision making \cite{morgenstern1953theory}. In a three-person game, a reciprocal relationship among agents could emerge and a coalition for which they take cooperative behavior helps them pursuit their own benefit. This leads to the concept of a cooperative game, in which players act in deference to an agreement. In our previous discussion in Sec. \ref{sec:entangled_goods}, when quantum strategies are not entangled, all players have a chance to increase their own profits by themselves, by exchanging quantum states appropriately. So the game is completely written in terms of a non-cooperative quantum game theory. In this section, we investigate an economy where quantum strategies are generally entangled. Our main interest here can be stated in a twofold way: 1) Can a quantum state become a medium of exchange? 2) Can a player find an optimal quantum strategy when strategies are entangled? In the previous section, we showed that a quantum state can be a medium of exchange in the sense that the associated classical term becomes a steady state. Again, readers will find the first question can be affirmatively solved (see Table \ref{tab:strategies}, for example). A remarkable result in this section is the existence of an optimal quantum strategy that allows Bob to obtain his consuming good $\ket{2}$ from Alice regardless of her inclination \eqref{eq:Bob_to_Alice}. Alice has no way to decline this trade with Bob and may be forced to have a state that she does not desire, hence she may suffer a loss. To compensate her loss, Alice responds to Bob's attack by counter-attacking, but it turns out that her action may cause damage not only to Bob but also to Charlie due to entanglement. So Bob and Charlie share a mutual interest and they have a motivation to form a coalition. Indeed Bob and Charlie can find an amicable solution that does not decrease their profits (Fig:\ref{fig:flow2}). This novel property only becomes evident after quantum goods and entangled strategies are introduced, and cannot be explained by conventional classical games or the corresponding quantum variants \cite{PhysRevLett.83.3077,PhysRevLett.82.1052,2001PhRvA..64c0301B}. Those salient features of our quantum extension of the Kiyotaki-Wright model strongly suggests that general quantum extensions of conventional games, including the contract theory and mechanism design, offers novel and various possibilities that cannot be discussed in the framework of the traditional economics. In other words, economics in the quantum era will become very different from economics today. This provides incentives to study quantum economics.

In the previous section, we consider a trade of entangled quantum goods, and we see the quantum entanglement gradually disappears through repetitions of a trade. Now we discuss the case of unentangled quantum goods and entangled quantum strategies. Following our previous discussions, we again employ $\mathcal{J}=\exp{\left[i \theta Y^{\otimes 12}\right]}$ for entanglement among strategies. Quantum strategies are maximally entangled when $\theta=\frac{\pi}{4}$, which we use throughout this section. We are interested in a game where a quantum strategy that do not commute $\mathcal{J}$ is used for a deal. Suppose Alice, Bob and Charlie choose the following strategies when $\frac{c_3-c_2}{\delta u}$ and $\frac{c_2-c_1}{\delta u}$ are in the red region of Fig.\ref{fig:optimal}
\begin{align}
\begin{aligned}
\label{eq:quantum_strategy}
U_A^2&=Y \otimes I & U_A^3&=\sqrt{1-q_A}Y\otimes Y+\sqrt{q_A}Y\otimes X \\
U_B^1&=Y \otimes I &U_B^3&=\sqrt{1-q_B}Y\otimes Y+\sqrt{q_B}Z\otimes Y \\
U_C^1&=I\otimes Y & U_C^2&=\sqrt{1-q_C}Y\otimes Y+\sqrt{q_C}Z\otimes Y 
\end{aligned}
\end{align}
By choosing $q_B=0$ Bob can reproduce his classical strategy ($s_B=1$ at \eqref{eq:ex}). An important difference from Sec.\ref{sec:entangled_goods} is that Bob may use quantum strategy $Z$ when Bob has $\Ket{3}$. If there is no entanglement, Bob's state does not change from $\ket{3}$ to $\Ket{1}$. In this case, however the strategy has other implications. Bob desires to exchange his state $\Ket{3}$ for Alice's $\Ket{2}$, by which Alice may suffers a loss. If Alice, Bob and Charlie have $\Ket{2},\Ket{3}$ and $\Ket{1}$, then their initial state $\Ket{231}$ is mapped to
\begin{align}
\begin{aligned}
\label{eq:Bob_to_Alice}
\Ket{231}&\Ket{W}\Ket{0000}_A\otimes\Ket{0000}_B\otimes\Ket{0000}_C \otimes\Ket{0}_A\Ket{0}_B\Ket{0}_C\\
\to&\frac{i\sqrt{q_B}}{\sqrt{3}}\Ket{331}\Ket{110}\Ket{0111}_A\otimes\Ket{1110}_B\otimes\Ket{1111}_C\otimes\Ket{0}_A\Ket{1}_B\Ket{0}_C\\
&-\frac{i\sqrt{1-q_B}}{\sqrt{3}}\Ket{231}\Ket{110}\Ket{1000}_A\otimes\Ket{0011}_B\otimes\Ket{0000}_C\otimes\Ket{0}_A\Ket{0}_B\Ket{0}_C\\
&-\frac{1}{\sqrt{3}}\Ket{231}\Ket{101} \Ket{1000}_A\otimes\Ket{0000}_B\otimes\Ket{0100}_C\otimes\Ket{0}_A\Ket{0}_B\Ket{0}_C\\
&+\frac{i\sqrt{q_B}}{\sqrt{3}}\Ket{211}\Ket{011}\Ket{1111}_A\otimes\Ket{1110}_B\otimes\Ket{1011}_C\otimes\Ket{0}_A\Ket{0}_B\Ket{1}_C \\
&-\frac{i\sqrt{1-q_B}}{\sqrt{3}}\Ket{211}\Ket{011}\Ket{0000}_A\otimes\Ket{0011}_B\otimes\Ket{0100}_C\otimes\Ket{0}_A\Ket{0}_B\Ket{1}_C .
\end{aligned}
\end{align}
It is important to note that $\ket{231}$ is mapped on to $\ket{311}$ if Alice and Bob meet. This implies that, regardless of Alice's strategy $q_A$, Bob creates a new $\ket{3}$ by exchanging his $\ket{3}$ for Alice's $\Ket{2}$. This happens since Bob uses a quantum strategy that does not commute with $\mathcal{J}$ and Alice and Bob are entangled. This is a key characteristic of our quantum game in the following two senses: 1) classical games does not allow those solutions 2) this occurs even for a single stage game. The first property is clearly important since it distinguishes a quantum game from a classical one. Alice receives $\ket{3}$ in a different logic that she accepts it in the classical speculative equilibrium. Apart from that, here we would like to emphasize the second characteristic. As explained in \cite{2002PhRvA..66b4306V,2018arXiv180307919S}, the EWL-types of single stage quantum games \cite{PhysRevLett.83.3077} are entirely non-quantum mechanical and there are no radical solutions created thus far. Therefore, to our best knowledge, our work is the first model that reaches a milestone for this problem. Note that the basic part of our quantum circuit (Fig. \ref{fig:two_persons}) is the EWL protocol for the prisoner's dilemma \cite{PhysRevLett.83.3077}. {References \cite{2018arXiv180307919S, PHOENIX2020126299} are standard reviews of the WEL protocol.} Quantum properties of the infinitely repeated quantum prisoner's dilemma are reported in \cite{ikeda_foundation_2019,2020arXiv200505588A}

Similarly we find $\ket{311}$ is mapped on to $\ket{231}$ if Alice and Bob meet \eqref{eq:311Q}. This implies that, regardless of Bob's strategy $q_B$, Alice creates a new $\ket{1}$ by exchanging her $\ket{3}$ for Bob's $\Ket{1}$.
\begin{align}
\begin{aligned}\label{eq:311Q}
\Ket{311}&\Ket{W}\Ket{0000}_A\otimes\Ket{0000}_B\otimes\Ket{0000}_C \otimes\Ket{0}_A\Ket{0}_B\Ket{0}_C\\
\to&-\frac{i\sqrt{q_A}}{\sqrt{3}}\Ket{231}\Ket{110}\Ket{1100}_A\otimes\Ket{0111}_B\otimes\Ket{1111}_C\otimes\Ket{1}_A\Ket{0}_B\Ket{0}_C\\
&-\frac{i\sqrt{1-q_A}}{\sqrt{3}}\Ket{311}\Ket{110}\Ket{0011}_A\otimes\Ket{1000}_B\otimes\Ket{0000}_C\otimes\Ket{0}_A\Ket{0}_B\Ket{0}_C\\
&-\frac{1}{\sqrt{3}}\Ket{311}\Ket{011} \Ket{0000}_A\otimes\Ket{1000}_B\otimes\Ket{0100}_C\otimes\Ket{0}_A\Ket{0}_B\Ket{0}_C\\
&-\frac{i\sqrt{q_A}}{\sqrt{3}}\Ket{211}\Ket{101}\Ket{1100}_A\otimes\Ket{1111}_B\otimes\Ket{1011}_C\otimes\Ket{1}_A\Ket{0}_B\Ket{1}_C \\
&-\frac{i\sqrt{1-q_A}}{\sqrt{3}}\Ket{211}\Ket{101}\Ket{0011}_A\otimes\Ket{0000}_B\otimes\Ket{0100}_C\otimes\Ket{1}_A\Ket{0}_B\Ket{1}_C 
\end{aligned}
\end{align}

The transition matrix $T$ corresponding to the strategies \eqref{eq:quantum_strategy} is 
\begin{align}
    T=
    \begin{pmatrix}
    \frac{2}{3} & 0 & \frac{1}{3} & 0 & \frac{1}{3} & \frac{1}{3} & 0 & 0 \\
0 & \frac{1}{3} & 0 & 0 & 0 & 0 & 0 & 0 \\
\frac{1}{3} & \frac{1}{3} & \frac{2-q_B}{3} & \frac{1}{3} & \frac{q_A}{3} & 0 & \frac{1}{3} & \frac{1}{3} \\
0 & \frac{1}{3} & 0 & \frac{2-q_B}{3} & 0 & \frac{1}{3} & 0 & 0 \\
0 & 0 & 0 & 0 & \frac{2-q_A}{3} & 0 & \frac{1}{3} & 0 \\
0 & 0 & 0 & 0 & 0 & 0 & 0 & 0 \\
0 & 0 & \frac{q_B}{3} & 0 & 0 & \frac{1}{3} & \frac{1}{3} & \frac{1}{3} \\
0 & 0 & 0 & \frac{q_B}{3} & 0 & 0 & 0 & \frac{1}{3} \\
    \end{pmatrix}.
\end{align}
Even if Charlie has a quantum good $\ket{2}$ initially, he loses it in the next step and never receives $\Ket{2}$ again. He eventually gets $\ket{1}$. So we do assume Charlie has the good $\Ket{2}$. Therefore the corresponding fixed point is 
\begin{align}
\label{eq:fixed_point}
    \rho=\frac{
    (2+2q_A+2q_B)\Ket{211} \bra{211}+
    (2q_A+2)\Ket{231} \bra{231}
    +q_B \Ket{311} \bra{311}
    +q_B(1+q_A)\Ket{331}\bra{331}}{4(1+q_A)+q_B(3+q_A)}.
\end{align}
So the state $\Ket{211}$ seems most likely to appear. Let us discuss equilibrium strategies. The expected total profit $V_i(i=A,B,C)$ \eqref{eq:expected_payoff} of Alice, Bob and Charlie can be expressed as 
\begin{align}
\begin{aligned}
    V_A&=\left[-(c_2,c_2,c_3,c_3)+\frac{u\delta}{3}(1,0,1+q_A,1) \right]\frac{1}{4(1+q_A)+q_B(3+q_A)}
        \begin{pmatrix}
    2+2q_A+q_B \\
    2+2q_A \\
    q_B \\
    q_B(1+q_A) \\
    \end{pmatrix} \\
        V_B&=\left[-(c_1,c_3,c_1,c_3)+\frac{u\delta}{3}(1,q_B,0,0) \right]\frac{1}{4(1+q_A)+q_B(3+q_A)}
        \begin{pmatrix}
    2+2q_A+q_B \\
    2+2q_A \\
    q_B \\
    q_B(1+q_A) \\
    \end{pmatrix} \\
    V_C&=\left[-(c_1,c_1,c_1,c_1)+\frac{u\delta}{3}(0,1,1,2) \right]\frac{1}{4(1+q_A)+q_B(3+q_A)}
        \begin{pmatrix}
    2+2q_A+q_B \\
    2+2q_A \\
    q_B \\
    q_B(1+q_A) \\
    \end{pmatrix},
\end{aligned}
\end{align}
where each vector is represented with the basis $\{\Ket{211}, \Ket{231},\Ket{311},\Ket{331}\}$, that constitute \eqref{eq:fixed_point}. Consulting \eqref{eq:V}, we find
\begin{align}
\begin{aligned}
    v_A&=(c_2,c_2,c_3,c_3) ,& w_A&=\frac{1}{3}(1,0,1+q_A,1) \\
    v_B&=(c_1,c_3,c_1,c_3),& w_B&=\frac{1}{3}(1,q_B,0,0) \\
    v_C&=(c_1,c_1,c_1,c_1),&w_C&=\frac{1}{3}(0,1,1,2)
\end{aligned}
\end{align}

\begin{figure}[H]
\centering
\includegraphics[width=\hsize,bb=0 0 1303 365]{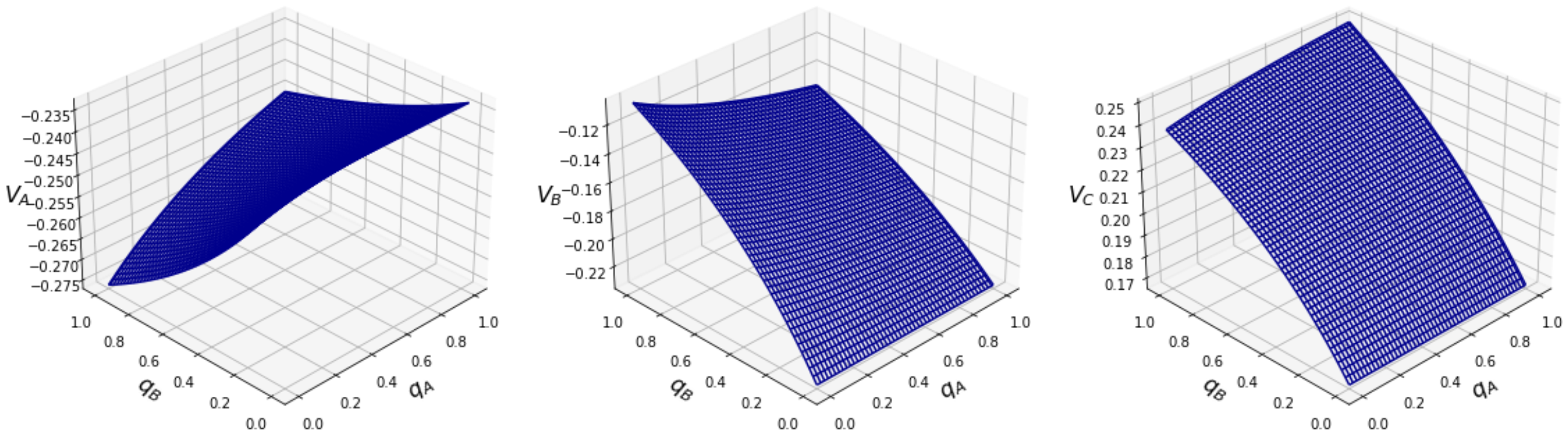}
\if{\begin{minipage}{0.32\hsize}
    \centering
     \includegraphics[width=\hsize, bb=0 0 375 342]{VA_quantum_at_(0.4,0.4).png}
\end{minipage}
\begin{minipage}{0.32\hsize}
    \centering
     \includegraphics[width=\hsize, bb=0 0 375 342]{VB_quantum_at_(0.4,0.4).png}
\end{minipage}
\begin{minipage}{0.32\hsize}
    \centering
    \includegraphics[width=\hsize, bb= 0 0 375 342]{VC_quantum_at_(0.4,0.4).png}
\end{minipage} }\fi
\caption{Total expected profit of Alice, Bob and Charlie when $\frac{c_2-c_1}{u\delta}=0.4,\frac{c_3-c_2}{u\delta}=0.4$.}
\label{fig:my_label}
\end{figure}

For any $q_B\in(0,1]$, Alice's optimal strategy which maximizes $V_A$ is $q_A=1$ since 
\begin{align}
    \frac{\partial V_A}{\partial q_A} =\frac{(c_3-c_2)q_B(4-q_B)+\delta u q_B^2}{(4(1+q_A)+q_B(3+q_A))^2} \geq 0.
\end{align}
The inequality saturates if and only if $q_B=0$. So any $q_A$ can be her optimal strategy when $q_B=0$. Similarly, for any $q_A\in[0,1]$, Bob's optimal strategy is also $q_B=1$ since 
\begin{align}
        \frac{\partial V_B}{\partial q_B} =\frac{(c_3-c_1)2(1-q_A)(1+q_A)+\delta u2(1+q_A)^2}{(4(1+q_A)+q_B(3+q_A))^2} >0
\end{align}
Therefore $q_A=q_B=1$ is an equilibrium. Alice gains more rewards compared with the case when she chooses the fundamental strategy ($q_A=0$). Bob and Charlie succeed in increasing their rewards. By substituting $q_A=q_B=1$ into \eqref{eq:fixed_point}, we find the corresponding steady state 
\begin{align}
\label{eq:steady_state4}
    \rho=\frac{5}{12} \Ket{211} \bra{211}+\frac{4}{12} \Ket{231} \bra{231}+\frac{1}{12} \Ket{311} \bra{311}
    +\frac{2}{12}\Ket{331}\bra{331}.
\end{align}

\begin{figure}[H]
    \centering
    \includegraphics[width=4cm, bb=0 0 100 100]{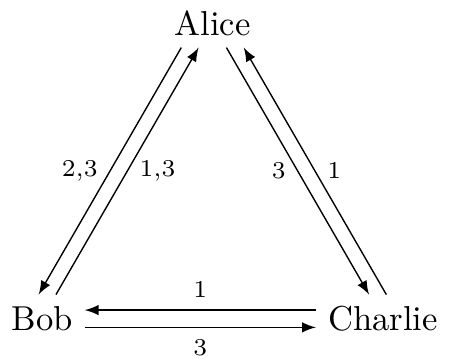}
    \caption{The flow of goods associated with the steady state \eqref{eq:steady_state4}. The difference from the classical case is that $\ket{3}$ is communicated between Alice and Bob, which happens when the game is in a state $\ket{231}$ or $\ket{311}$.}
    \label{fig:my_label}
\end{figure}
Bob's strategy $U_B^3=Z \otimes Y$ decreases Alice's profit. As her counter strategy, she may use $U_A^2=X\otimes I$ when she has $\Ket{2}$, in order to prevent Bob from exchanging his $\ket{3}$ for her $\Ket{2}$. In addition, if Charlie plays $U_C^1=I\otimes Y$, then Alice can receive $\Ket{1}$ and increase her profit by exchanging her $\Ket{2}$ for $\Ket{1}$. Quantum strategies and the flow of quantum goods are summarized in Table.\ref{tab:strategies}. It is important that there are several strategies that make $\Ket{1}$, $\Ket{2}$ or $\Ket{3}$ media of exchange.
\begin{figure}[H]
    \centering
    \includegraphics[width=\hsize,bb=0 0 1500 1665]{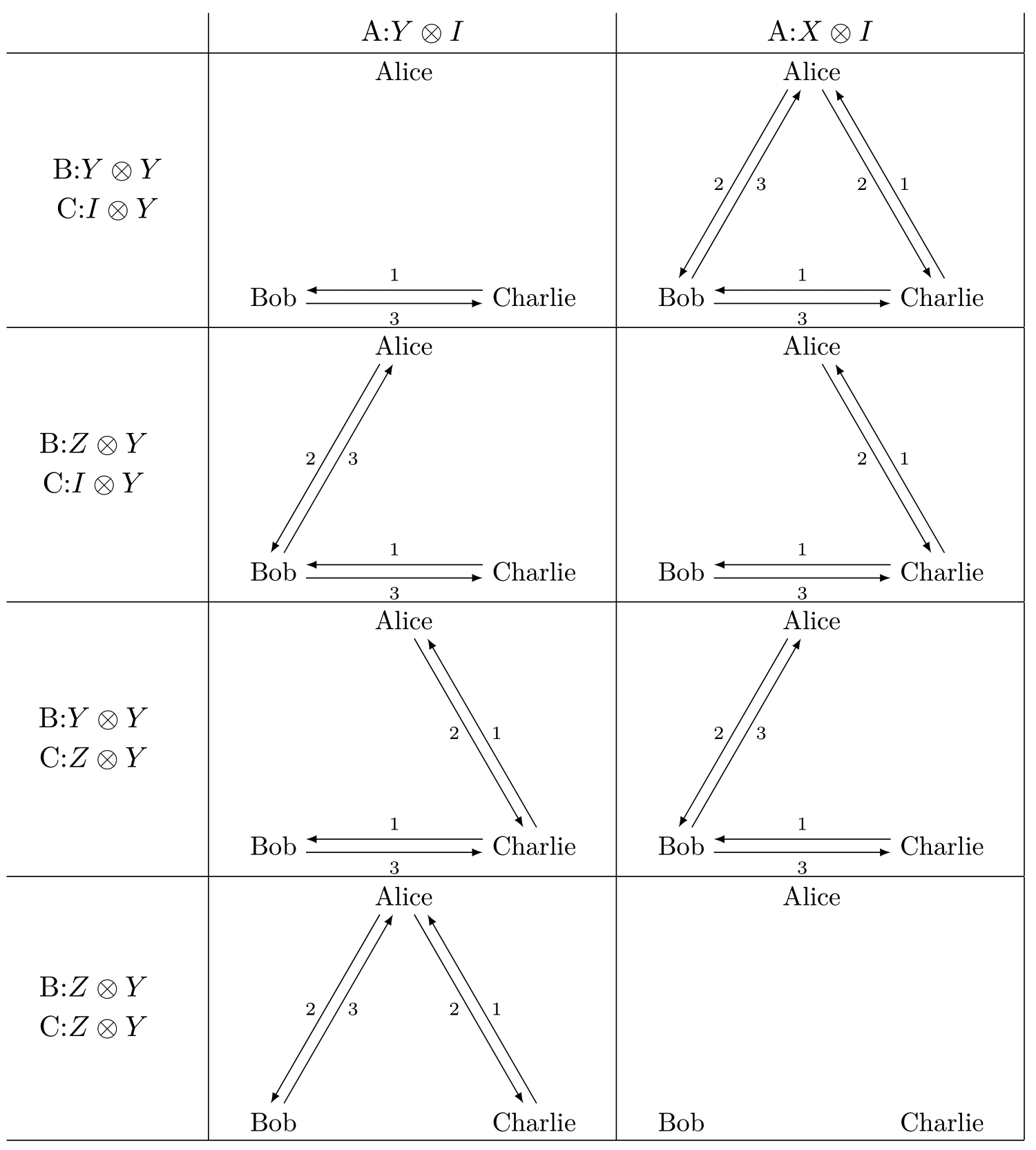}
\caption{Quantum strategies and the flow of quantum goods when $\Ket{231}$ is an initial state of Alice, Bob and Charlie who are allowed to decide $U^2_A$, $U^3_B$ and $U^1_C$ independently. If Alice uses $X\otimes I$, she and Charlie can exchange $\Ket{2}\leftrightarrow \Ket{1}$ and she prevents Bob from using his previous strategy $Z\otimes Y$ for exchanging $\Ket{2}\leftrightarrow \Ket{3}$. However if Bob uses $Y\otimes Y$, the trade $\Ket{2}\leftrightarrow \Ket{3}$ can be realized. Nothing is mediated in the bottom right case.}
    \label{tab:strategies}
\end{figure}

\if{\begin{table}[H]
\centering
\resizebox{0.70\columnwidth}{!}{%
    \begin{tabular}[c]{c|c|c|} 
    &A:$Y\otimes I$ & A:$X\otimes I$\\ \hline
\begin{tabular}{c}
             B:$Y\otimes Y$  \\
             C:$I\otimes Y$ \\
        \end{tabular} &\begin{minipage}{5truecm}
      \centering
      \includegraphics[width=2cm, bb=0 0 100 100]{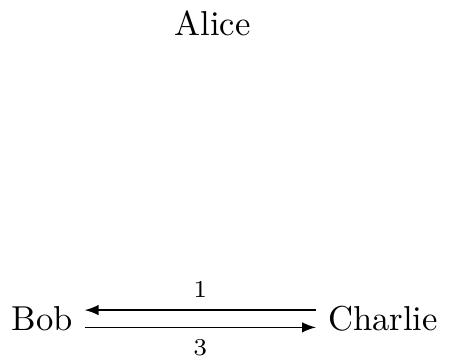}
    \end{minipage}
        &     \begin{minipage}{5truecm}
        \centering
    \includegraphics[width=2cm, bb=0 0 100 100]{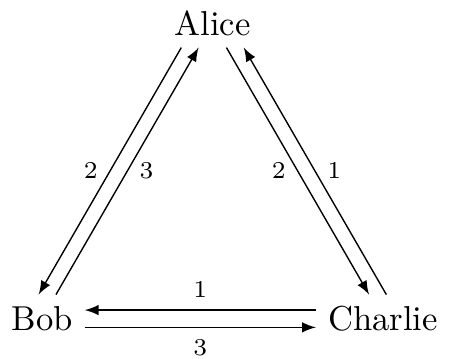}
    \end{minipage} \\ \hline
\begin{tabular}{c}
            B:$Z\otimes Y$  \\
            C:$I\otimes Y$ \\
        \end{tabular}& 
        \begin{minipage}{5truecm}
        \centering
    \includegraphics[width=2cm, bb=0 0 100 100]{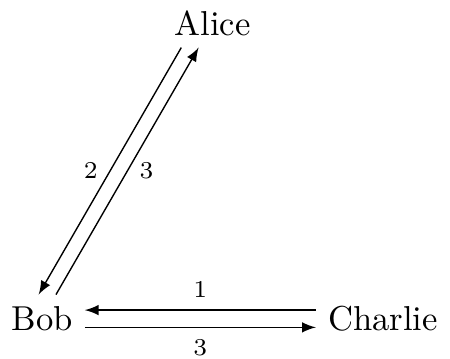}
    \end{minipage}
    &\begin{minipage}{5truecm}
      \centering
      \includegraphics[width=2cm, bb=0 0 100 100]{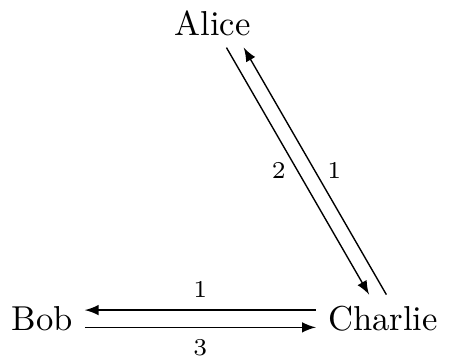}
    \end{minipage}
    \\ \hline
\begin{tabular}{c}
             B:$Y\otimes Y$  \\
             C:$Z\otimes Y$ \\
        \end{tabular}&
        \begin{minipage}{5truecm}
      \centering
      \includegraphics[width=2.5cm, bb=0 0 100 100]{auto2_3.pdf}
    \end{minipage}
    &\begin{minipage}{5truecm}
      \centering
      \includegraphics[width=2.5cm, bb=0 0 100 100]{auto2_2.pdf}
    \end{minipage}
    \\ \hline
\begin{tabular}{c}
             B:$Z\otimes Y$  \\
             C:$Z\otimes Y$ \\
        \end{tabular}&
        \begin{minipage}{5truecm}
      \centering
      \includegraphics[width=2.5cm, bb=0 0 100 100]{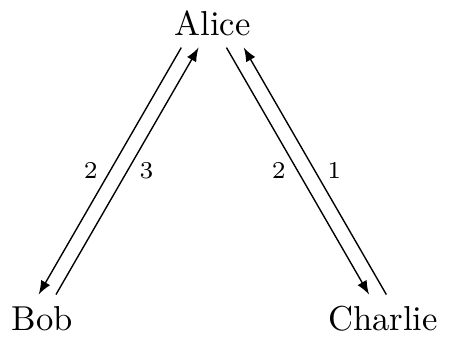}
    \end{minipage}
    &\begin{minipage}{5truecm}
      \centering
      \includegraphics[width=2.5cm, bb=0 0 100 100]{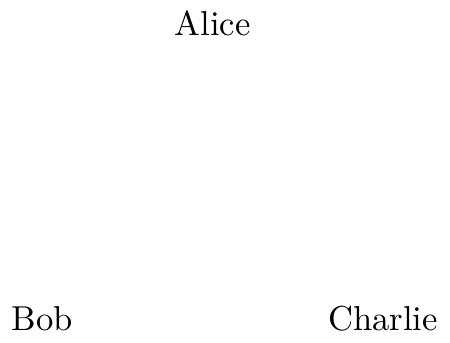}
    \end{minipage}
    \\ \hline
\end{tabular}}
\caption{Quantum strategies and the flow of quantum goods when $\Ket{231}$ is an initial state of Alice, Bob and Charlie who are allowed to decide $U^2_A$, $U^3_B$ and $U^1_C$ independently. If Alice uses $X\otimes I$, she and Charlie can exchange $\Ket{2}\leftrightarrow \Ket{1}$ and she prevents Bob from using his previous strategy $Z\otimes Y$ for exchanging $\Ket{2}\leftrightarrow \Ket{3}$. However if Bob uses $Y\otimes Y$, the trade $\Ket{2}\leftrightarrow \Ket{3}$ can be realized. Nothing is mediated in the bottom right case.}
    \label{tab:strategies}
\end{table} }\fi

As Charlie gets his consuming good, a trade $\ket{1}\leftrightarrow \ket{3}$ between Bob and Charlie is the best trade for Charlie and, as Bob reduces his storage cost, it is also a good trade for Bob. So they may want to trade. However if Bob and Charlie use $Z\otimes Y$ and $Z\otimes Y$ respectively, they do not reach a mutually acceptable agreement on the trade $\ket{1}\leftrightarrow \ket{3}$. To solve the problem, they cooperate and try to increase their profits by means of the following strategy. Suppose both of Bob and Charlie choose the strategy \eqref{eq:1} with the probability $p$ and choose the strategy \eqref{eq:2} with the probability $1-p$. 
\begin{align}
\label{eq:1}
&\left\{
\begin{array}{l}
U_B^1=Y\otimes I  \\
U_C^1=I\otimes Y 
\end{array}
\right.&& 
\begin{array}{l}
U_B^3=\sqrt{1-q_B}Y\otimes Y+\sqrt{q_B}Z\otimes Y \\
U_C^2=Y\otimes Y
\end{array}
\\
\label{eq:2}
&\left\{
\begin{array}{l}
U_B^1=Y\otimes I   \\
U_C^1=\sqrt{1-q^\prime_C}I\otimes Y+\sqrt{q^\prime_C}Z\otimes Y 
\end{array}
\right.&&
\begin{array}{l}
U_B^3=Y\otimes Y \\
U_C^2=Y\otimes Y
\end{array}
\end{align}

Bob and Charlie make an arrangement and decide a single $p\in[0,1]$. Alice may want to use 
\begin{align}
    U_A^2&=\sqrt{1-q^\prime_A} Y\otimes I+\sqrt{q^\prime_A}X\otimes I &
    U_A^3&=Y\otimes X 
\end{align}
If Alice uses $q_A^\prime=0$, Bob and Charlie can maximize their profits by choosing $p=1, q_B=1$. Similarly, if she uses $q_A^\prime=1$, they can maximize their profits by choosing  $p=0,q_C=1$. Let $T_1$ and $T_2$ be the transition matrices associated with strategies \eqref{eq:1} and \eqref{eq:2}. The total transition matrix is $T=pT_1+(1-p)T_2$. Their explicit matrix representations are  
\begin{align}
    T_1=
    \begin{pmatrix}
    \frac{2-q_A^\prime}{3}  & 0 &  \frac{1}{3}  & 0 &  \frac{1}{3}  &  \frac{1}{3}  & 0 & 0 \\
\frac{q_A^\prime}{3}  &  \frac{1}{3}  & 0 & 0 & 0 & 0 & 0 & 0 \\
\frac{1}{3}  &  \frac{1}{3}  &  \frac{2-q_B(1-q_A^\prime)-(1-q_B)q_A^\prime-q_A^\prime}{3}  &  \frac{1}{3}  &  \frac{1}{3}  & 0 &  \frac{1}{3}  &  \frac{1}{3}  \\
0 &  \frac{1}{3}  & \frac{q_A^\prime}{3} &  \frac{2-q_B(1-q_A^\prime)-(1-q_B)q_A^\prime}{3}  & 0 &  \frac{1}{3}  & 0 & 0 \\
0 & 0 & 0 & 0 &  \frac{1}{3}  & 0 &  \frac{1}{3}  & 0 \\
0 & 0 & 0 & 0 & 0 & 0 & 0 & 0 \\
0 & 0 &  \frac{q_B(1-q_A^\prime)+(1-q_B)q_A^\prime}{3}  & 0 & 0 &  \frac{1}{3}  &  \frac{1}{3}  &  \frac{1}{3}  \\
0 & 0 & 0 &  \frac{q_B(1-q_A^\prime)+(1-q_B)q_A ^\prime}{3}  & 0 & 0 & 0 &  \frac{1}{3}  \\
    \end{pmatrix} \\
    T_2=
    \begin{pmatrix}
        \frac{2-q_A^\prime(1-q_C^\prime)-(1-q_A^\prime)q_C^\prime}{3}  & 0 &  \frac{1}{3}  & 0 &  \frac{1}{3}  &  \frac{1}{3}  & 0 & 0 \\
\frac{q_A^\prime(1-q_C^\prime)+(1-q_A^\prime)q_C^\prime}{3}  &  \frac{1}{3}  & 0 & 0 & 0 & 0 & 0 & 0 \\
\frac{1}{3}  &  \frac{1}{3}  &  \frac{2-q_A^\prime-q_A^\prime(1-q_C^\prime)-(1-q_A^\prime)q_C^\prime}{3}  &  \frac{1}{3}  &  \frac{1}{3}  & 0 &  \frac{1}{3}  &  \frac{1}{3}  \\
0 &  \frac{1}{3}  & \frac{q_A^\prime(1-q_C^\prime)+(1-q_A^\prime)q_C^\prime}{3} &  \frac{2-q_A^\prime}{3}  & 0 &  \frac{1}{3}  & 0 & 0 \\
0 & 0 & 0 & 0 &  \frac{1}{3}  & 0 &  \frac{1}{3}  & 0 \\
0 & 0 & 0 & 0 & 0 & 0 & 0 & 0 \\
0 & 0 &  \frac{q_A^\prime}{3}  & 0 & 0 &  \frac{1}{3}  &  \frac{1}{3}  &  \frac{1}{3}  \\
0 & 0 & 0 &  \frac{q_A^\prime}{3}  & 0 & 0 & 0 &  \frac{1}{3}  \\
    \end{pmatrix}
\end{align}

It is only the pair of Bob and Charlie who can establish the coalition as described. Both of them can increase their profit without loss. This is not true for Alice and Bob, or for Alice and Charlie. If Alice and Bob exchange $\ket{2}\leftrightarrow\Ket{3}$, then Bob gains but Alice suffers a loss. Similarly if Alice and Charlie trade $\ket{2}\leftrightarrow\Ket{1}$, then Alice benefits but Charlie incurs a loss. The emergence of such a coalition is based on a typical quantum phenomena that has not been reported in the past studies on the Kiyotaki-Write model. The flow of quantum commodities under those strategies are summarized in Fig. \ref{fig:flow2}.   

\section{Conclusion and Discussion}
The main contributions of our article are summarized as follows. In Sec. \ref{sec:money}, we introduced our quantum game model which describes the emergence of quantum commodity money as an equilibrium strategy of quantum exchange. In Sec. \ref{sec:circuit} we presented quantum circuits for consensus building and exchanging quantum states. In Sec. \ref{sec:entangled_goods} we addressed generic entangled states and clarified the mechanism of the quantum economy leading a quantum state to a medium of exchange. In Sec. \ref{sec:entangled_strategy} we studied some equilibria of entangled quantum strategies and demonstrated how a coalition can emerge from a non-cooperative game.

We propose several research directions for readers interested in quantum games and quantum economics. First of all, it is straightforward but important to explore an economy that allows quantum fiat money to be a medium of exchange. For this purpose one should simply add $\Ket{4}:=\Ket{00}$ as a quantum object with no intrinsic value. Moreover it will be a good exercise to implement our algorithms (Fig.\ref{fig:two_persons} and Fig.\ref{fig:three_persons}) with a quantum computer or a quantum simulator to study a quantum algorithm of consensus building and verify our results numerically. In addition, it is also interesting to study the quantum model of the type B economy, where Alice, Bob and Charlie produce $\Ket{3}$, $\Ket{1}$ and $\Ket{2}$, respectively. As shown in the original Kiyotaki-Wright model \cite{10.2307/1832197}, the dominant equilibrium strategy depends on models and it will be also true for our quantum extension. Indeed, as discussed in Sec. \ref{sec:entangled_strategy}, a choice of initial states is crucial for the optimal strategies. Moreover it is exciting to address generic entangled strategies. In our analysis we focused on the simplest case $\mathcal{J}=\exp\left[i \theta Y^{\otimes 12}\right]$. Our three-person quantum game with generic entangled strategies is very complicated and there are many open questions left. Furthermore, it is also possible to analyse equilibrium strategies of our economy where quantum walkers \cite{gudder2014quantum,PhysRevA.48.1687,1996JSP....85..551M} enjoy business. In this article, we used $\Ket{W}$ to decrease complexity, but it is yet another important question whether a quantum state can be a medium of exchange even if players encounter in a different way. For this purpose, it may be helpful to use the following matrices to introduce quantum walk
\begin{equation}
    \begin{pmatrix}
    a_{11}&a_{12}&a_{13}\\
    0&0&0\\
    0&0&0
    \end{pmatrix},~
    \begin{pmatrix}
    0&0&0\\
    a_{21}&a_{22}&a_{23}\\
    0&0&0
    \end{pmatrix},~
    \begin{pmatrix}
    0&0&0\\
    0&0&0\\
    a_{31}&a_{32}&a_{33}
    \end{pmatrix}
\end{equation}
where $|a_{i1}|^2+|a_{i2}|^2+|a_{i3}|^2=1$ for $i=1,2,3$. 

For advanced readers, it is meaningful to investigate a general quantum theory of economics, including market design and contracts \cite{cat}, so that we can provide efficient quantum information and communication markets in the upcoming era of quantum technology. For example, one can refer to our algorithm to consider many-to-many matching. Moreover, our study suggests that entangled strategies case a situation that is not predetermined in an agreement, thereby it will be complex and costly for the parties to make their contract complete. That means quantum theory also sheds new light on incomplete contracts \cite{grossman1986costs,10.2307/1912698,maskin1999unforeseen}. Quantum theory of economics will also shed new light on macroeconomics. In particular, we guess different countries will start using different quantum currencies. It is an open problem to study how quantum currencies circulate in the global economy. A formal model for a classical setup is provided in \cite{10.2307/2298058}. The more quantum technologies develop, the more important quantum economical perspectives will become. In addition, it will contribute to development of legal systems in the quantum era, since economics are inextricable from jurisprudence.  

\section*{Acknowledgement}
We thank Avi Beracha for carefully reading the manuscript. This work was supported by PIMS Postdoctoral Fellowship Award (K. I.). 


\section*{Author Contribution}
K. I. designed and performed the research, interpreted the results, and wrote the paper. S. A. helped with the calculations.
\bibliographystyle{utphys}
\bibliography{ref}
\end{document}